\documentclass[english,superscriptaddress,showpacs,manuscript]{revtex4-1}
\usepackage[T1]{fontenc}
\usepackage[latin9]{inputenc}
\usepackage[active]{srcltx}
\usepackage{mathtools}
\usepackage{bm}
\usepackage{amsmath}
\usepackage{amssymb}
\usepackage{graphicx}
\usepackage{esint}
\usepackage{color}

\makeatletter

\newcommand{\lyxmathsym}[1]{\ifmmode\begingroup\def\b@ld{bold}
  \text{\ifx\math@version\b@ld\bfseries\fi#1}\endgroup\else#1\fi}

\@ifundefined{textcolor}{}
{%
 \definecolor{BLACK}{gray}{0}
 \definecolor{WHITE}{gray}{1}
 \definecolor{RED}{rgb}{1,0,0}
 \definecolor{GREEN}{rgb}{0,1,0}
 \definecolor{BLUE}{rgb}{0,0,1}
 \definecolor{CYAN}{cmyk}{1,0,0,0}
 \definecolor{MAGENTA}{cmyk}{0,1,0,0}
 \definecolor{YELLOW}{cmyk}{0,0,1,0}
}

\makeatother

\usepackage{babel}
\begin{document}
\global\long\def\vc#1{\bm{#1}}

\title{Electron scattering and transport in liquid argon}

\author{G. J. Boyle}

\affiliation{College of Science, Technology \& Engineering, James Cook University,
Townsville 4810, Australia}

\author{R. P. McEachran}

\affiliation{Research School of Physical Sciences and Engineering, Australian
National University, Canberra, ACT 0200, Australia}

\author{D. G. Cocks}

\affiliation{College of Science, Technology \& Engineering, James Cook University,
Townsville 4810, Australia}

\author{R. D. White}

\affiliation{College of Science, Technology \& Engineering, James Cook University,
Townsville 4810, Australia}
\begin{abstract}
The transport of excess electrons in liquid argon driven out of equilibrium
by an applied electric field is revisited using a multi-term solution
of Boltzmann's equation together with ab initio liquid phase cross-sections
calculated using the Dirac-Fock scattering equations. The calculation
of liquid phase cross-sections extends previous treatments to consider
multipole polarisabilities and a non-local treatment of exchange while
the accuracy of the electron-argon potential is validated through
comparison of the calculated gas phase cross-sections with experiment.
The results presented highlight the inadequacy of local treatments
of exchange that are commonly used in liquid and cluster phase cross-section
calculations. The multi-term Boltzmann equation framework accounting
for coherent scattering enables the inclusion of the full anisotropy
in the differential cross-section arising from the interaction and
the structure factor, without an a priori assumption of quasi-isotropy
in the velocity distribution function. The model, which contains no
free parameters and accounts for both coherent scattering and liquid
phase screening effects, was found to reproduce well the experimental
drift velocities and characteristic energies.
\end{abstract}
\maketitle

\section{Introduction}

The study of electron transport in non-polar liquids is of fundamental
interest for understanding the dynamics of electronic processes in
liquids and disordered systems, including dynamic and scattering processes.
More recently, attention has focussed on applications including liquid
state electronics, driven by use in high-energy particle detectors
such as the liquid argon time projection chamber (LArTPC). Advances
in the fields of plasma discharges in liquids and associated electrical
breakdowns (see e.g. the review of Bruggmen \cite{Bruggeman2009})
are dependent on a fundamental knowledge of charged particle transport
in liquids. Furthermore, the rapidly developing interdisciplinary
field of plasma medicine requires \cite{Kong2009,Tian2014,Norberg2014,Chen2014}
a detailed knowledge of electron transport through liquid water and
other biostructures, typically under non-equilibrium conditions.

The study of excess electrons in dense gases and fluids is a complex
problem, requiring the inclusion of many effects that are not present
in dilute gaseous systems. The major contributions to these effects
arise from the small interparticle spacings and their highly correlated
separations. For thermal energies, the de Broglie wavelengths of the
excess electrons are often orders of magnitude larger than the interatomic
spacing, which leads to significant quantum-like effects. Even within
a semi-classical picture, where the excess electrons are assumed to
act as point-like particles, no particular volume is ``owned'' by
a single atom. This means the typical picture for transport in a gas,
i.e. a series of individual collision events separated by the mean-free
path, is no longer valid, making it important to consider multiple
scattering effects of the electron from many atoms simultaneously.
Of particular note is the effect of ``coherent scattering'' and
the pair correlations of the liquid play a very important role in
this and other effects.

Many previous calculations for electrons in dense systems have neglected
these liquid effects for simplicity, modelling dense fluids by applying
a theory for dilute gases with only an appropriate increase of the
density. However, a few alternative theories exist that have explored
liquids in different ways.  Borghesani et al~\cite{Borghesani2006}
have heuristically combined the liquid effects identified above to
obtain an effective cross-section. When used in the standard equations
from kinetic theory for mobility in a non-zero field, their results
have been shown to be remarkably accurate.  Braglia and Dallacasa
\cite{Braglia82} have derived a theory that addresses both enhancements
and reductions to the zero-field mobility through a Green's function
approach with appropriate approximations to the self-energy but do
not go beyond linear response theory and hence do not explain non-equilibrium
behaviour at high fields.

In contrast to the above approaches, the seminal articles by Lekner
and Cohen \cite{Cohen1967,Lekner1967} outline a method to address
effects of a dense fluid from an ab initio approach by appropriate
modifications of the microscopic processes. The article by Lekner
\cite{Lekner1967} describes how an effective potential for a single
collision event can be built up from knowledge of only the single-atom/electron
potential and the pair correlator of the fluid as well as prescribing
a method for obtaining effective cross-sections from this potential.
The article by Cohen and Lekner \cite{Cohen1967} then describes how
the effects of coherent scattering can be included with these effective
cross-sections in a Boltzmann equation solution for the calculation
of transport properties. Sakai et al \cite{Sakai2007a} have been
able to improve agreement with experiment by empirically modifying
the resultant cross-sections of the Cohen and Lekner formalism and
by including inelastic processes. Atrazhev et al \cite{Atrazhev1981}
were able to simplify the arguments of Lekner \cite{Lekner1967} to
argue that, for small energies, the effective cross-section becomes
dependent on the density only and obtained good agreement with experiment.
However the distance at which to enforce this new behaviour of the
effective cross-section remains a free parameter in the theory and
this constant effective cross-section must be found empirically. Atrazhev
and co-workers went on to consider the interaction as a muffin tin
potential, with each cell being a Wigner-Seitz sphere surrounding
each atom in the liquid. They used a variable phase function method
which could describe the absence of a Ramsauer minimum in the liquid
cross-section along with density fluctuations of the liquid \cite{Atrazhev1995,Atrazhev1996,Atrazhev1998}.

The calculations we present in this paper are based on a generalization
of the Cohen and Lekner formalism, overcoming several approximations
which are no longer necessary in modern day transport and scattering
theory. With regard to the scattering potential, Lekner \cite{Lekner1967}
used the Buckingham polarization potential \cite{Buckingham38} as input, which
we will show is completely inadequate due to its omission of the exchange
interaction. This is not noticeable for gas phase measurements, due
to the fitting parameter of the Buckingham potential, but shows significant
differences after the liquid modifications are applied. By performing
a detatiled analysis of the partial phase shifts, Atrazhev and co-workers
\cite{Atrazhev1996} were able to isolate the important properties
of the potential which are required for accurate determination of
the transport properties. Our calculations instead avoid these difficulties
by using accurate forms for the electron-atom interaction.

With regard to the transport theory itself, we employ a previously
derived extension of the Cohen and Lekner formalism for the Boltzmann
equation from a two-term to a full multiterm treatment of the velocity
distribution function \cite{White2011}. This theory utlises the full
anisotropic detail of the cross-sections that is available in our
calculations. For dilute gaseous systems, the two-term approximation
can be in serious error \cite{White2003a}, and in this study we consider
contributions to the error arising from the neglect of the full anisotropy
in both the velocity distribution function and the differential scattering
cross-section for liquid systems. We perform calculations specifically
for the noble gas of argon, which is an excellent test bed for new
theories due to the good availability of experimental data and the
high degree of accuracy to which ab initio calculations can model
the gaseous phase. Available experimental data include drift velocities
and characteristic energies in both the gas and liquid phases, as
well as precise single-atom cross-sections.  We emphasize that we
are interested in the full non-equilibrium description of the transport
properties and not only that of zero-field mobilities, and so we must
consider the full range of the static structure factor $S(K)$ instead
of $S(0)$ which is fixed by the isothermal compressibility.

In the following sections we consider the calculation of the macroscopic
swarm transport properties in the gaseous and liquid environments
from the microscopic cross-sections, modified by the screening and
coherent scattering effects discussed above. We first detail the calculation
of the gas phase cross-sections in section \ref{sec:gas_scattering},
using accurate potentials in the Dirac-Fock scattering equations and
then address, in section \ref{sec:Screening-of-the}, effects of screening
in the liquid. The transition from a gas to liquid requires a modification
of the scattering to include an effective scattering potential and
an effective non-local exchange term which we describe in section
\ref{sec:Effective-potential-in}. The application of these cross
sections in structured media is outlined in section \ref{TransportTheory-1}
and we present the results of our transport calculations in section
\ref{sec:Results}. Initially in Section \ref{sub:Electrons-in-gaseous}
we consider only the gas phase, understanding the importance of an
accurate treatment of exchange and polarization and thereby establishing
the credibility of the initial gas-phase potential subsequently used
as input for the calculation of cross-sections for the liquid phase
environment. Transport coefficients calculated using the screened
cross-sections and associated coherent scattering effects are considered
in Section \ref{sub:Electrons-in-liquid}, where they are compared
with the available measured transport data. Throughout this paper
we will make use of atomic units ($m=e=a_{0}=\hbar=1$) unless otherwise
specified.

\section{Scattering of electrons by Argon gas}

\label{sec:gas_scattering}The core of a transport calculation is
based on an accurate description of the scattering of the electron
off a particle in the bulk. Effective interaction potentials are often
used to determine various measurable properties, such as scattering
lengths or polarizabilities. These effective potentials are successful
so long as they correctly reproduce these quantities for input in
other simulations. However, as mentioned above, there are many additional
effects due to a dense gas or liquid which can modify the details
of the scattering processes. Hence, we require a potential that does
not only produce the correct scattering properties in the dilute limit
but also well describes the scattering properties under a perturbation
of the potential.

\global\long\def\d{\mathrm{d}}
\global\long\def\np{{\phantom{.}}}
 \global\long\def\rp{{\rm p}}
 \global\long\def\rs{{\rm s}}
 \global\long\def\barc{\bar{c}}
 \global\long\def\barl{\bar{l}}
 \global\long\def\barv{\bar{v}}
 \global\long\def\jhat{\hat{\jmath}}
 \global\long\def\nhat{\hat{n}}
 \global\long\def\rshalf{\mbox{\ensuremath{{\scriptstyle \frac{1}{2}}}}}
 \global\long\def\scleb#1#2#3#4#5{C(#1\,#2\,#3;#4\,#5)}
 \global\long\def\sclebsq#1#2#3#4#5{C^{2}(#1\,#2\,#3;#4\,#5)}
 \global\long\def\linewpq#1#2{\overline{W}_{\!\!#1}^{\,#2}}
 \global\long\def\lineW{{\overline{W}}}

In the pure elastic energy region, in addition to the static potential, there 
are only two interactions which need to be taken into account in electron-atom 
collisions, namely polarization and exchange. The polarization can be accounted 
for by means of long-range multipole polarization potentials while the exchange 
interaction is represented most accurately by a short-range non-local potential 
formed by antisymmetrizing the total scattering wavefunction. 

In the present work the scattering of the incident electrons, with
wavenumber $k$, by argon atoms is described in the gaseous phase
by the integral formulation of the partial wave Dirac-Fock scattering
equations (see \cite{Chen08} for details). In matrix form, these
equations can be written as 
\begin{align}
\begin{pmatrix}f_{\kappa}(r)\\
g_{\kappa}(r)
\end{pmatrix}=\begin{pmatrix}v_{1}(kr)\\
v_{2}(kr)
\end{pmatrix}+\frac{1}{k}\int_{0}^{r}\!\!\!\d x\, G(r,x)\,\biggl[U(x)\begin{pmatrix}f_{\kappa}(x)\\
g_{\kappa}(x)
\end{pmatrix}-\begin{pmatrix}\lineW_{Q}(\kappa;x)\\
\lineW_{P}(\kappa;x)
\end{pmatrix}\biggr] & \;,\label{eq:dirac_eqn}
\end{align}
where the local potential $U(r)$ is given by the sum of the static
and local polarization potentials i.e., 
\begin{align}
U(r)=U_{\rs}(r)+U_{\rp}(r)\label{eq:local_potential}
\end{align}
and $\lineW_{P}(\kappa;r)$ and $\lineW_{Q}(\kappa;r)$ represent
the large and small components of the exchange interaction. In equation~\eqref{eq:dirac_eqn},
$f_{\kappa}(r)$ and $g_{\kappa}(r)$ are the large and small components
of the scattering wavefunction where the quantum number $\kappa$
can be expressed in terms of the total and orbital angular momentum
quantum numbers $j$ and $l$ according to 
\begin{align}
j=\vert\kappa\vert-\frac{1}{2}\quad\mbox{with}\quad l=\begin{cases}
\kappa, & \text{if \ensuremath{\kappa>0}}\\
-\kappa-1, & \text{if \ensuremath{\kappa<0}}
\end{cases} & \;.
\end{align}

The free particle Green's function $G(r,x)$ in equation~\eqref{eq:dirac_eqn} is 
defined in terms of Riccati-Bessel and Riccati-Neumann functions (see 
equations~(23) and~(24a,b) of ref.~\cite{Chen08}). The kinetic energy $\epsilon$ 
of the incident electron and its wavenumber
$k$ are related by 
\begin{align}
k^{2}=\frac{1}{\hbar^{2}c^{2}}\,\epsilon\,\bigl(\epsilon+2mc^{2}\bigr) & \;.
\end{align}
We note that if we ignore $\epsilon$ with respect to $2mc^{2}$,
we obtain the usual non-relativistic relationship between the wavenumber
and the energy of the incident electron. 

The static potential $U_{\mathrm{s}}(r)$ in equation~\eqref{eq:local_potential}
is determined in the usual manner from the Dirac-Fock orbitals of
the atom \cite{Chen08}. The polarization potential $U_{\rp}(r)$
was determined using the polarized orbital method \cite{McEachran77}
and contained several static multipole terms as well as the corresponding
dynamic polarization terms \cite{McEachran90,Mimnagh93}. In total,
the potential $U_{\rp}(r)$ contained all terms up to and including
those that behave as $r^{-14}$ asymptotically.

Finally, the exchange terms $\lineW_{P}(\kappa;r)$ and $\lineW_{Q}(\kappa;r)$
in equation~\eqref{eq:dirac_eqn} are given by
\begin{align}
\lineW_{P\;{\rm or}\; Q}(\kappa_{2};r) & =(1+\gamma)\sum_{n'\kappa'}\bigl\{ 
P_{n'\kappa'}(r)\;{\rm or}\; 
Q_{n'\kappa'}(r)\bigr\}\bigl\{-\bigl[\epsilon_{n'\kappa'}+\epsilon\bigr]\,\Delta_{n'\kappa'}\,\delta(\kappa,\kappa')\,\nonumber 
	\\
 & 
	\quad+e^{2}\sum_{\nu}q_{n'\kappa'}\,\frac{1}{2\nu+1}\,\sclebsq{j}{j'}{\nu}{-\rshalf}{\rshalf}\frac{1}{r}\,y_{\nu}(n'\kappa',\kappa;r)\bigr\} 
	\label{eq:exchange_terms}
\end{align}
Here, $\scleb j{j'}{\nu}{-\rshalf}{\rshalf}$ is the usual Clebsch-Gordan
coefficient and the sum over $n'\kappa'$ in 
equation~\eqref{eq:exchange_terms}
is over the radial part of the atomic orbitals ($P_{n'\kappa'}(r))$ and 
$Q(n'\kappa')(r)$) of the
ground state while $q_{n'\kappa'}=2j'+1$ is the occupation number
of these closed sub-shells where the $\epsilon_{n'\kappa'}$ are the
eigenvalues of these sub-shells. The exact form of the definite integral 
$\Delta_{n'\kappa'}$ and the indefinite integral 
$r^{-1}\,y_\nu(n'\kappa',\kappa;r)$ are given in equations~(11) and~(12) of 
ref.  ~\cite{Dorn98}.

We note that the dependence of the exchange terms \eqref{eq:exchange_terms}
on the wave function requires an iterative solution for equation \eqref{eq:dirac_eqn}.

In the integral equation formulation, the scattering phase shifts
can be determined from the asymptotic form of the large component
of the scattering wavefunction i.e., 
\begin{align}
f_{\kappa}(r)\underset{{\scriptstyle r\rightarrow\infty}}{\longrightarrow}A_{\kappa}\,\jhat_{l}(kr)-B_{\kappa}\,\nhat_{l}(kr) & \;,
\end{align}
where 
\begin{align}
A_{\kappa}=1-\frac{1}{k}\int_{0}^{\infty}\!\!\d r\,\Bigl\{\barv_{1}(kr)\,\bigl[U(r)\, f_{\kappa}(r)-\lineW_{P}(\kappa;r)\bigr]+\barv_{2}(kr)\,\bigl[U(r)\, g_{\kappa}(r)-\lineW_{Q}(\kappa;r)\bigr]\Bigr\}\label{eq:Ak_integral}
\end{align}
and 
\begin{align}
B_{\kappa}=-\frac{1}{k}\int_{0}^{\infty}\!\!\d r\,\Bigl\{ v_{1}(kr)\,\bigl[U(r)\, f_{\kappa}(r)-\lineW_{P}(\kappa;r)\bigr]+v_{2}(kr)\,\bigl[U(r)\, g_{\kappa}(r)-\lineW_{Q}(\kappa;r)\bigr]\Bigr\} & \;.\label{eq:Bk_integral}
\end{align}
The partial wave phase shifts are then given by 
\begin{align}
\tan\delta_{l}^{\pm}(k)=\frac{B_{\kappa}}{A_{\kappa}} & \;,
\end{align}
where the $\delta_{l}^{\pm}$ are the spin-up (+) and spin-down $(-)$
phase shifts. 

The total elastic and momentum transfer cross-sections are given,
in terms of these phase shifts, by 
\begin{align}
\sigma_{{\rm el}}(k^{2})=\frac{4\pi}{k^{2}}\sum_{l=0}^{\infty}\Bigl\{(l+1)\,\sin^{2}\delta_{l}^{+}(k)+l\,\sin^{2}\delta_{l}^{-}(k)\Bigr\}\label{eq:el_cs_def}
\end{align}
and 
\begin{align}
\sigma_{{\rm mt}}(k^{2}) & =\frac{4\pi}{k^{2}}\sum_{l=0}^{\infty}\Bigl\{\frac{(l+1)(l+2)}{2l+3}\,\sin^{2}\bigl(\delta_{l}^{+}(k)-\delta_{l+1}^{+}(k)\bigr)+\frac{l(l+1)}{2l+1}\,\sin^{2}\bigl(\delta_{l}^{-}(k)-\delta_{l+1}^{-}(k)\bigr)\\
 & \quad+\frac{(l+1)}{(2l+1)(2l+3)}\,\sin^{2}\bigl(\delta_{l}^{+}((k)-\delta_{l+1}^{-}(k)\bigr)\Bigr\}\nonumber 
\end{align}
which can be shown to reduce to the non-relativistic results if we
set $\delta_{l}^{+}(k)=\delta_{l}^{-}(k)=\delta_{l}(k)$. 

\begin{figure}
\includegraphics[width=0.6\textwidth]{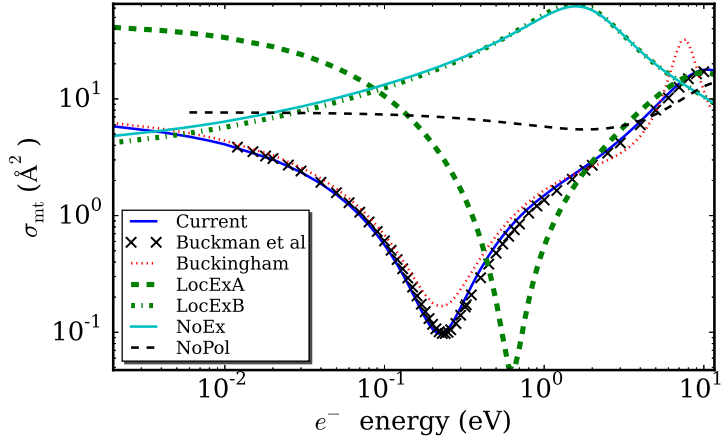}

\protect\caption{\label{fig:cross_sections}Cross-sections for electron scattering
from argon. The calculations described in this paper, which use the
non-local exchange interaction (solid line) are in good agreement
with the recommended set of Buckman et al. \cite{Buckman03}. A comparison
with a calculation similar to that of Lekner \cite{Lekner1967} using
a Buckingham potential (dotted line) shows loose qualitative agreement
at the Ramsauer minimum, but quantitatively is incorrect. Also shown
are the results from using two different effective models of a local
exchange potential (thick dashed \cite{Furness73} and dash-dotted
lines \cite{Bransden76}) which do not agree with experimental measurement
at all, as well as the cross-section when exchange is included but
polarization is neglected (thin dashed line).}
\end{figure}

As can be seen in Figure~\ref{fig:cross_sections}, neither the polarization
nor the exchange interaction alone is capable of reproducing the true
Ramsauer minimum in the argon momentum transfer cross-section; it
is only when we combine these two interactions that there is agreement
with experiment. This is also true for the Ramsauer minimum in the
elastic cross-section.

In the original work of \cite{Lekner1967}, Lekner described the elastic
scattering of electrons by argon atoms by just the local dipole polarization
potential of Buckingham \cite{Buckingham38} which is given by 
\begin{align}
U_{\rp}(r)=-\frac{\alpha_{\d}}{2\,(r^{2}+r_{a}^{2})^{2}} & \;,\label{eq:Buckingham_pot}
\end{align}
where $\alpha_{\d}$ is the static dipole polarizability of argon
and $r_{a}$ is an adjustable parameter. Lekner chose this parameter
so as to obtain the experimental scattering length %
\mbox{%
$a_{0}=-1.5$~a.u.%
} of~\cite{Frost64}. This value is very close to the current recommended
value of %
\mbox{%
$a_{0}=-1.45$~a.u.%
} of \cite{Buckman03}. The value obtained in the current work is %
\mbox{%
$a_{0}=-1.46$~a.u.%
} 

As a consequence of Lekner's choice for the adjustable parameter $r_{a}$,
his simple polarization potential in equation~\eqref{eq:Buckingham_pot}
was able to mimic the effects of both the polarization and exchange
interactions at low energies of the incident electron and his calculation
was able to produce a low-energy Ramsauer minimum in the momentum
transfer cross-section. At higher energies his momentum transfer cross-section
deviates from the experimental cross-section.

We show our cross-sections calculated using (\ref{eq:el_cs_def})
in Figure~\ref{fig:cross_sections}. We obtain very good agreement
with the recommended set of cross-sections of \cite{Buckman03} which
combine many different experimental measurements and theoretical calculations.
In order to demonstrate the importance of including the non-local
exchange interaction, we have also repeated the calculation using
two different model potentials that replace the non-local exchange
with an effective local term in the potential \cite{Furness73,Bransden76}.
One of these local approximations \cite{Furness73} is qualitatively
wrong, showing the same behaviour as that without exchange. The other
approximation \cite{Bransden76} is qualitatively similar but differs
in the scattering length and position of the Ramsauer minimum by an
order of magnitude. It is clear to see that there is a significant
difference between the results. When we compare our results to those
of the Buckingham potential, where we set $r_{\alpha}=1.087$~a.u.
such that the scattering length is $a_{0}=-1.50$~a.u., we find that
it does follow the general shape of the Ramsauer minimum. However,
we emphasize that this is a result of the fitting parameter $r_{\alpha}$
and this potential does not accurately describe the details of the
scattering.

\section{Screening of the polarization interaction\label{sec:Screening-of-the}}

The effects of the high density of the liquid are included in our
calculations by several modifications of the gas scattering properties.
The first of these is to account for the screening of a single induced
atomic dipole by the induced dipoles of all other atoms. Our procedure
outlined in this section closely follows that of Lekner \cite{Lekner1967}.

In the dilute gas limit, the mobile electron undergoes a collision
with a single atom of the gas effectively in isolation from all other
atoms in the gas. During this collision the electron induces a set
of multipole moments in the atom, which in turn interact with the
electron through a charge-multipole potential, resulting in the polarization
potential, $U_{\mathrm{p}}(r)$, of section \ref{sec:gas_scattering}
above. For a dilute gas, the range of the potential produced by these
induced multipole moments is relatively small compared with the large
interatomic spacing and so it is a good approximation to neglect their
effect on other atoms. However, with higher densities of the gas or
liquid, many atoms can have a non-negligible induced set of multipole
moments originating from both the mobile electron and from all other
atoms in the bulk. The effective charge-multipole polarization potential
felt by the electron at any particular location $\bm{r}_{e}$ is then
the sum of the polarization potentials from all atoms.

We consider effects originating from the induced dipoles of the atoms
only and determine the effective polarization of an individual atom
self-consistently. We first assume that the induced dipole strength
for every atom in the bulk can be written as $f(r)\alpha_{d}(r)e/r^{2}$
where $r$ is the distance of the electron from the atom, $\alpha_{d}(r)$
is the exterior dipole polarizability (see \cite{McEachran77b}, equation
(1)) for a single atom that results from the interaction with the
electron, and $f(r)$ accounts for polarization screening which must
be determined. This simple multiplicative factor is valid, so long
as we average over the atomic distribution. In the dilute-gas limit,
we can safely approximate $f(r)=1$, and in the dense case we must
obtain a self-consistent expression for $f(r)$. By choosing a particular
``focus atom'' $i$ at location $\bm{r}_{i}$ such that $\bm{r}=\bm{r}_{e}-\bm{r}_{i}$,
and assuming that the coefficient $f(r)$ is known for all other atoms,
which we denote by $f_{\mathrm{bulk}}(r)$, we can calculate \cite{Lekner1967}
the dipole strength for atom $i$ from:
\begin{equation}
f_{i}(r)=1-\pi N\int_{0}^{\infty}ds\:\frac{g(s)}{s^{2}}\int_{|r-s|}^{r+s}dt\:\Theta(r,s,t)\frac{\alpha_{d}(t)f_{\mathrm{bulk}}(t)}{t^{2}}\label{eq:dipole_selfcon}
\end{equation}
which has been obtained using bipolar coordinates, $s$ and $t$,
where $N$ is the density of the bulk, $g(s)$ is the isotropic pair
correlator of the bulk and the factor 
\begin{equation}
\Theta(r,s,t)=\frac{3}{2}\frac{(s^{2}+t^{2}-r^{2})(s^{2}+r^{2}-t^{2})}{s^{2}}+(r^{2}+t^{2}-s^{2}),
\end{equation}
arises due to the form of the electric field of a dipole. The integrations
over $s$ and $t$ represent the contribution from an atom located
at a distance $s$ from atom $i$ and a distance $t$ from the electron.
The likelihood of finding an atom is determined by $g(s)$ and so
it can be seen that equation (\ref{eq:dipole_selfcon}) approximates
the exact polarization by that resulting from the ensemble average
of all atomic configurations, given that one atom is located at $\bm{r}_{i}$.
In this approximation, the polarization itself is always aligned along
the vector $\hat{\bm{r}}$ between the focus atom and the electron.

\begin{figure}
\includegraphics[width=0.6\textwidth]{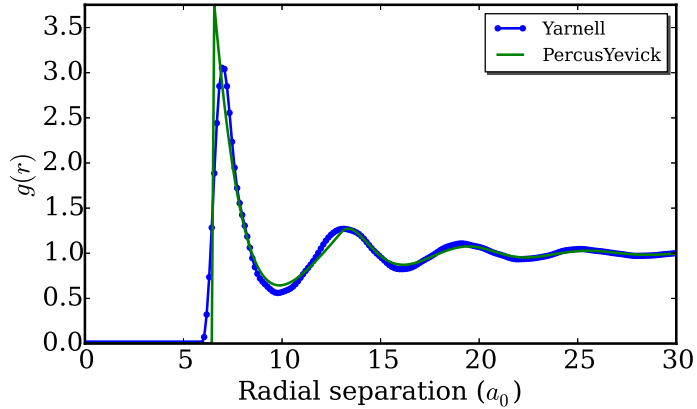}

\protect\caption{\label{fig:pair_correlator}Pair correlator for argon, as reported
in Yarnell \cite{Yarnell73}, measured in neutron scattering experiments.
Also plotted, is the pair correlator calculated in the analytical
Percus-Yevick approximation as used by Lekner \cite{Lekner1967}.}
\end{figure}

The self-consistent solution to equation (\ref{eq:dipole_selfcon})
is obtained by setting $f_{i}(r)=f_{\mathrm{bulk}}(r)$ and solving
for $f_{i}(r)$, which we do by iteration. The most important quantity
in equation (\ref{eq:dipole_selfcon}) is the pair correlator, which
represents the next order in the particle distribution in the bulk
beyond the average density. In the calculation of Lekner, the pair
correlator was taken to be the analytical solution of the Percus-Yevick
model for ease of calculation. In our calculation, we go beyond this
by using the experimental measurements of Yarnell \cite{Yarnell73}
to more accurately describe the correlations. The data we use, which
was obtained for a bulk density of $N=0.0213\,\textrm{\AA}^{-3}$,
is shown in Figure~\ref{fig:pair_correlator} and compared with the
Percus-Yevick model at the same density. 

\begin{figure}
\includegraphics[width=0.6\textwidth]{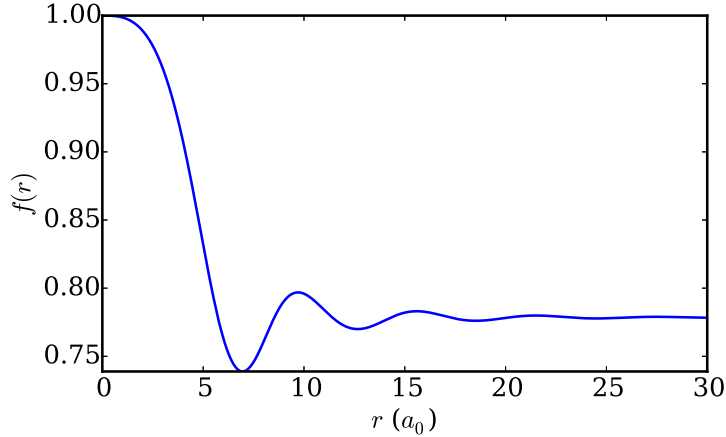}

\protect\caption{\label{fig:screening_func}The screening function $f(r)$ of the polarization
interaction potential for scattering of an electron from a single
argon atom in a bulk of density %
\mbox{%
$N=0.0213\,\textrm{\AA}^{-3}$%
}.}

\end{figure}

Using the pair correlator of argon, we have self-consistently calculated
the screening function $f(r)$ and show the result in Figure~\ref{fig:screening_func}.
Although this screening factor technically applies to the dipole term
only, we work with a rather more complicated form of the polarization
term than Lekner had originally considered. However, as the largest
contribution to the polarization does indeed come from the dipole
term, we have decided to apply the screening factor $f(r)$ to the
entire polarization potential. Hence, with the screening of the polarization
taken into account, the screened polarization potential, $\tilde{U}_{p}(r)$,
of an electron with one atom in a dense fluid is given by:
\begin{align}
\tilde{U}_{p}(r) & =f(r)U_{\mathrm{p}}(r)\;.
\end{align}
We note that, in contrast to Lekner, who used only the static dipole
polarizability $\alpha_{d}$, the more accurate representation of
the atom-electron interaction as described in section \ref{sec:gas_scattering}
has already led to a radial dependence of the polarization potential
$U_{p}(r)$ beyond that of a potential whose asymptotic behaviour
is $r^{-4}$. The effect of the screening has hence led to a further
modification of $U_{p}(r)$ which is density dependent.

\section{Effective potential in liquid\label{sec:Effective-potential-in}}

For input into the kinetic theory, we require appropriate cross-sections
for the scattering of the electron from a single ``focus atom''
in the bulk. As discussed above, the presence of the other atoms screens
the polarization interaction between the electron and the focus atom.
However, there is another more obvious effect resulting from the other
atoms in the bulk: their interaction with the electron itself remains
significant even when the electron is very close to the focus atom.
Hence, as outlined in Lekner \cite{Lekner1967}, we build up an effective
potential that is experienced by the electron throughout a single
scattering event, as well as define what is meant by ``a single scattering
event''. Although we follow the general principles of \cite{Lekner1967},
we calculate the cross-sections in a distinctly different fashion.

The effective potential that we consider $U_{\mathrm{eff}}=U_{1}+U_{2}$
is made of two parts: $U_{1}(r)$ which corresponds to the direct
interactions with the focus atom, and $U_{2}(r)$, which corresponds
to the interaction of the electron with the rest of the bulk. As it
is prohibitively expensive to treat exact configurations of atoms
in the bulk, we build the external potential $U_{2}$ by again taking
the ensemble average:
\begin{equation}
U_{2}(r)=\frac{2\pi N}{r}\int_{0}^{\infty}dt\: U_{1}(t)\int_{|r-t|}^{r+t}ds\: sg(s)\;,\label{eq:U2}
\end{equation}
where the order of integration has been reversed in comparison to
(\ref{eq:dipole_selfcon}) for numerical convenience %
\footnote{In the arrangement of (\ref{eq:U2}), it is possible to precompute
the innermost integral cumulatively once and use its values in a look-up
table.%
}. We note that taking the ensemble average has the advantage of enforcing
spherical symmetry of the total effective potential $U_{\mathrm{eff}}$.
In calculating (\ref{eq:dipole_selfcon}) and (\ref{eq:U2}), we make
use of the quantity $\sigma_{\mathrm{core}}$, which corresponds to
the hard-core exclusion diameter for the distribution of atoms in
the bulk, i.e. the probability for two atoms to approach within a
distance $\sigma_{\mathrm{core}}$ is vanishingly small. For argon
$\sigma_{\mathrm{core}}\approx6\, a_{0}$ and we take advantage of
this by explicitly setting $g(s)=0$ for $s<\sigma_{\mathrm{core}}$
and adjusting the limits of equations (\ref{eq:dipole_selfcon}) and
(\ref{eq:U2}) accordingly.

In addition we go beyond Lekner's calculation by including the effects
of the exchange terms in the bulk. We do this by performing the same
ensemble average as in (\ref{eq:U2}) but over the quantities $\lineW_{P}$
and $\lineW_{Q}$ instead of $U_{1}$, obtaining bulk averages $\lineW_{P,2}$
and $\lineW_{Q,2}$. These are then included as effective exchange
terms, $\lineW_{(P\text{ or }Q),\mathrm{eff}}=\lineW_{P\text{ or }Q}+\lineW_{(P\text{ or }Q),2}$
in the Dirac-Fock scattering equations (\ref{eq:dirac_eqn}). In contrast
to $U_{2}$, these exchange terms are dependent on the wave function
itself, so the ensemble averages must be recalculated at every iteration
in the solution of (\ref{eq:dirac_eqn}).

\begin{figure}
\includegraphics[width=0.6\textwidth]{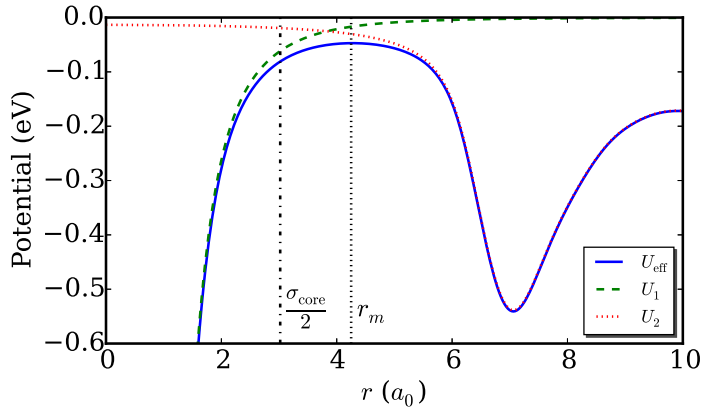}\protect\caption{\label{fig:potentials}Plots of the total effective potential $U_{\mathrm{eff}}$
felt by an electron when colliding with one atom in the liquid. Also
shown are the components, $U_{1}$ and $U_{2}$, which represent the
direct potential of the atom and the contribution of the remaining
atoms in the bulk respectively. The dashed vertical lines at $\sigma_{\mathrm{core}}/2$
and $r_{m}$ indicate the hard-core exclusion radius and the proposed
collisional sphere respectively. Note that effects of exchange are
not represented in this figure.}
\end{figure}

A plot of the functions $U_{\mathrm{eff}}$, $U_{1}$ and $U_{2}$
is shown in Figure~\ref{fig:potentials}. It can be seen that there
is a turning point that occurs at a distance we denote by $r_{m}$.
In the dense gas limit that we are investigating, this value is $r_{m}\approx4.3\, a_{0}$.
The turning point at $r_{m}$ provides a natural distinction between
the volume that is under the influence of the focus atom, i.e. the
sphere of radius $r_{m}$, and that of the rest of the bulk. Hence,
we can say that a single collision event takes place when an electron
enters and leaves the radius $r_{m}$ of a single atom. We note that
$r_{m}\approx\frac{2}{3}\sigma_{\mathrm{core}}>\sigma_{\mathrm{core}}/2$,
i.e. $r_{m}$ is larger than half of the minimal interatomic separation,
which could be considered to define the volume ``owned by'' the
focus atom and hence a logical choice for the ``collision event radius''.
$r_{m}$ is also different from the Wigner-Seitz diameter $d_{\mathrm{WS}}=2(4\pi N/3)^{-1/3}\approx4.2\, a_{0}$
\cite{Atrazhev1998}, although it is very similar.

We would now like to solve for the scattering properties, in particular
the cross-sections, from such a collision process. We assume that
it remains reasonable to extract the cross-sections through the phase
shifts in a partial wave expansion. In order to determine these, Lekner
chose to shift the effective potential by an amount $U_{0}$ such
that $U_{\mathrm{eff}}(r_{m})+U_{0}=0$, and to set the potential
$U_{\mathrm{eff}}(r>r_{m})=0$, and finally matched to the asymptotic
form of each partial wave in the usual fashion. In contrast, we choose
to leave the potential unaltered, but calculate the phase shift at
the point $r_{m}$ instead, effectively setting the upper limits of
equations (\ref{eq:Ak_integral}) and (\ref{eq:Bk_integral}) to be
$r_{m}$ instead of infinity. We note that this is also known as calculating
the ``phase function'' \cite{Atrazhev1996} at the point $r_{m}$,
which is equivalent to setting $U_{\mathrm{eff}}(r>r_{m})=0$ and
matching to the asymptotic form of the wave function. We believe that
this more accurately represents the available energy states in the
bulk. We denote this cross-section, including external contributions and 
screening effects, by $\sigma^\mathrm{scr}(\epsilon,\chi)$.

As we may assume $g(s)=0$ for $s<\sigma_{\mathrm{core}}$ and because
we calculate the potential only up to a distance of $r_{m}\approx\frac{2}{3}\sigma_{\mathrm{core}}$,
we can see that the integral over $t$ in (\ref{eq:U2}) is non-zero
only for $t\gtrsim\frac{1}{3}\sigma_{\mathrm{core}}\approx2\, a_{0}$.
At these ranges, the dominant contribution to the potential comes
from the polarization component. We also note that the values of $\lineW_{P,2}$
and $\lineW_{Q,2}$ are not well behaved for larger distances and
so we set them to be zero for $r>\sigma_{\mathrm{core}}/2$. We have
performed calculations that neglect the contribution of $\lineW_{P,2}$
and $\lineW_{Q,2}$ to the bulk and compared these to the full calculations,
which showed very little difference in the high energy regime of the
resultant cross-sections and a small difference of up to 5\% otherwise.
The effect of this change on the transport properties was a small
but non-negligible deviation.

\subsection{Cross-sections and variation of $r_{m}$}

\begin{figure}
\includegraphics[width=0.6\textwidth]{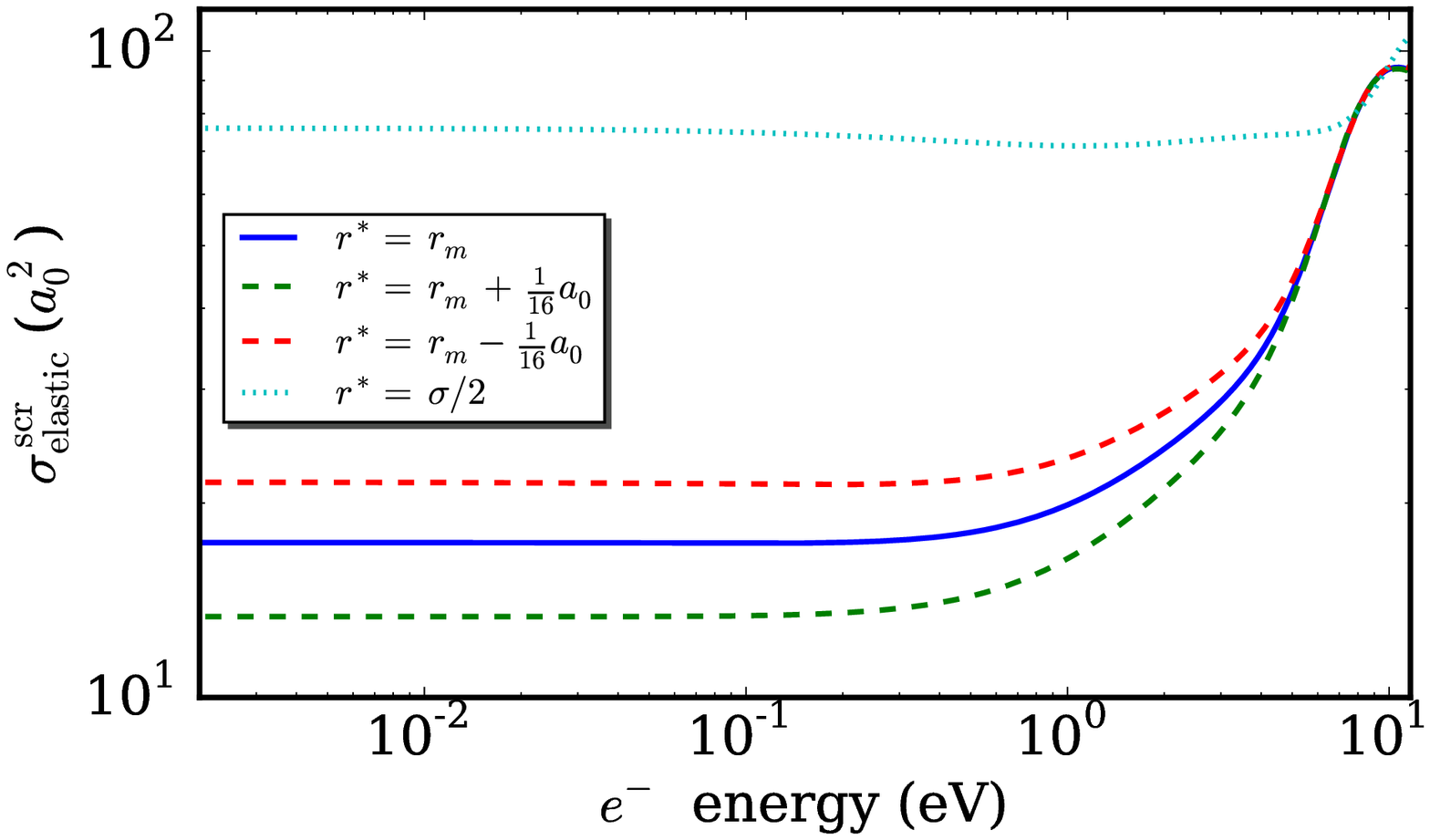}

\includegraphics[width=0.6\textwidth]{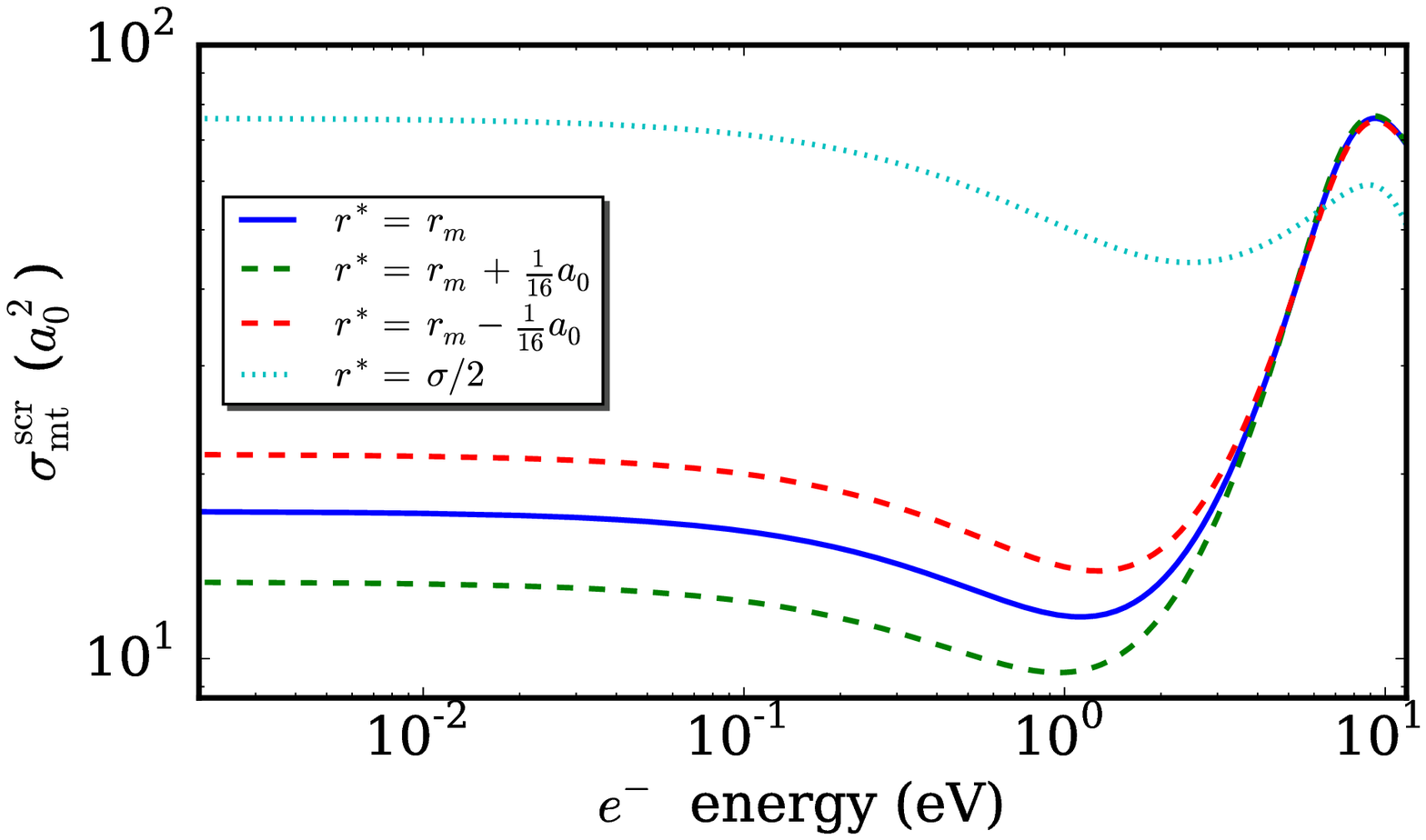}\protect\caption{\label{fig:Rm_variation}Screened 
	elastic total and momentum-transfer cross-sections
for argon calculated from the phase shifts determined at a distance
$r^{*}$. Our preferred choice for transport calculations in this
paper, $r^{*}=r_{m}$, corresponds to the solid line, the dashed lines
are those corresponding to a variations $r^{*}=r_{m}\pm\frac{1}{16}a_{0}$
and the dotted line corresponds to a variation of $r^{*}=\sigma_{\mathrm{core}}/2$.}
\end{figure}

The choice of the value for $r_{m}$ is a crucial part of our calculation.
It is worth mentioning that the choice we make above is consistent
in the limit of $N\rightarrow0$; in this case $U_{2}$ is so weak
that it is only after $U_{1}$ has significantly decayed for very
large values of $r$ that $d(U_{1}+U_{2})/dr=0$. Hence, $r_{m}\rightarrow\infty$
as $N\rightarrow0$ and our calculation reduces to the usual scattering
calculation from a single atom. However, in the dense case, it is
not known whether $d(U_{1}+U_{2})/dr|_{r_{m}}=0$ is the best choice
to model the scattering in the liquid. Hence, we have also performed
a sensitivity analysis on the parameter $r_{m}$. We denote the distance
at which we calculate the phase shifts by $r^{*}$ and allow it to
vary from our initial choice of $r^{*}=r_{m}$. The resultant cross-sections
from a variation of $\pm\frac{1}{16}a_{0}$ are shown in Figure~\ref{fig:Rm_variation}
as well as the more straightforward choice of $r^{*}=\sigma_{\mathrm{core}}/2$.
We note that Atrazhev et al \cite{Atrazhev1998} have implicitly investigated
this variation previously, in order to describe the effect of density
fluctuations on the effective cross-sections. In their case, the value
of $r^{*}$ was set to be the Wigner-Seitz cell radius, which itself
depends on the density of the liquid. In contrast, we keep the density
fixed while varying $r^{*}$.

It can easily be seen that the largest modification to the cross-sections
due to the variation in $r_{m}$ occurs at low energies. Importantly,
the more obvious choice of %
\mbox{%
$r^{*}=\sigma_{\mathrm{core}}/2$%
} yields a dramatically different behaviour. As will be shown later,
the effect that these variations have on the transport measurements
is significant and shifts the peak observed in various transport properties.

We note that we neglect the effect of density fluctuations, which
would modify the effective cross-section for the liquid. This was
investigated by Atrazhev et al \cite{Atrazhev1998}, and shown to
have a significant contribution to the cross-sections. However, their
article focused on a density for which the effective liquid cross-section
vanishes, causing the density fluctuations to be the largest contribution
for small electron energies. In our case, we can expect density fluctuations
to cause both enhancements and reductions of the cross-sections, which
would cancel out on average.

\section{Kinetic theory and transport properties\label{TransportTheory-1}}

\subsection{Multi-term solution of Boltzmann's equation}

The behaviour of electrons in gaseous and liquid argon, driven out
of equilibrium via an electric field $E$, can be described by the
solution of the Boltzmann's equation for the phase-space distribution
function $f(\vc r,\vc v,t)$~\cite{Boltzmann1872}: 

\begin{equation}
\frac{\partial f}{\partial t}+\vc v\cdot\nabla f+\frac{e\vc E}{m_{e}}\cdot\frac{\partial f}{\partial\vc v}=-J(f),\label{eq:BE}
\end{equation}
where $\vc r,$ $\vc v$ and $e$ denote the position, velocity and
charge of the electron respectively. The collision operator $J(f)$
accounts for interactions between the electrons of mass $m_{e}$ and
the background material. We restrict our considerations in this study,
to those applied reduced electric fields $E/N$ (where $N$ is the
number density of the background material) such that no internal states
of the individual argon atoms are excited.

To calculate the drift and diffusion coefficients, we represent the spatial dependence of the distribution functions as
\cite{Kumar1980d,Robson1986f}:
\begin{equation}
f\left(\vc r,\boldsymbol{v},t\right)=F(\vc v,t)n(\vc r,t)-F^{(L)}(\vc v)\frac{\partial n}{\partial z}-F^{(T)}(\vc v)\left[\cos\phi\frac{\partial n}{\partial x}+\sin\phi\frac{\partial n}{\partial y}\right]+...,
\end{equation}
where the superscripts $L$ and $T$ define quantities that are parallel
and transverse to the electric field (defined to be in the $z$ direction)
respectively. Solution of Boltzmann's equation~(\ref{eq:BE}) requires
decomposition of the coefficients in velocity space through an expansion in (associated) Legendre
polynomials: 
\begin{align}
F(\vc v)&=\sum_{l=0}^{\infty} F_{l}(v)P_{l}(\cos\theta)\; \nonumber\\
F^{(T)}(\vc v)&=\sum_{l=0}^{\infty} F_{l}^{(T)}(v)P_{l}^{1}(\cos\theta)\;,\label{sphericalharmonic-1}
\end{align}
%
%
%\begin{equation}
%F^{(L)}(\vc v)=\sum_{l=0}^{\infty}\, F_{l}^{(L)}(v)P_{l}(\cos\theta)\;,\label{sphericalharmonic-1-1}
%\end{equation}
where $\theta$ denotes the angle relative to the electric field direction
(taken to be the $z$-axis). 
%Likewise, the transverse component must
%be represented in terms of an expansion in associated Legendre polynomials,
%$P_{l}^{1}(\cos\theta)$:
%\begin{equation}
%F^{(T)}(\vc v)=\sum_{l=0}^{\infty}\, F_{l}^{(T)}(v)P_{l}^{1}(\cos\theta)\;.\label{sphericalharmonic-1-1-1}
%\end{equation}
This is a true multi-term solution of Boltzmann's equation, whereby
the upper bound in each of the $l$-summations are truncated at a
value $l_{\mathrm{max}}$, and this value is incremented until some
convergence criteria is met on the distribution function or its velocity
moments. By setting $l_{\mathrm{max}}=1$ we obtain the two-term approximation
commonly used in all electron transport theory in liquids \cite{Cohen1967,Borghesani2006,Sakai2007a},
which enforces a quasi-isotropic distribution. The current theory
does not make the quasi-isotropic assumption for the velocity distribution
function \textit{a priori}. By using the orthogonality of (associated)
Legendre polynomials, the following hierarchy of equations must be
solved to calculate the drift velocity and diffusion tensor \cite{Robson1986f}:
\begin{equation}
J^{l}F_{l}+\frac{l+1}{2l+3}a\left(\frac{\partial}{\partial v}+\frac{l+2}{v}\right)F_{l+1}+\frac{l}{2l-1}a\left(\frac{\partial}{\partial v}-\frac{l-1}{v}\right)F_{l-1}=0\label{eq:homogenous}
\end{equation}
%
%
%\begin{equation}
%J^{l}F_{l}^{(L)}+\frac{l+1}{2l+3}a\left(\frac{\partial}{\partial v}+\frac{l+2}{v}\right)F_{l+1}^{(L)}+\frac{l}{2l-1}a\left(\frac{\partial}{\partial v}-\frac{l-1}{v}\right)F_{l-1}^{(L)}=v\left(\frac{l+1}{2l+3}F_{l+1}+\frac{l}{2l-1}F_{l-1}\right)\label{eq:longitudinal}
%\end{equation}
%
\begin{equation}
J^{l}F_{l}^{(T)}+\frac{l+2}{2l+3}a\left(\frac{\partial}{\partial v}+\frac{l+2}{v}\right)F_{l+1}^{(T)}+\frac{l-1}{2l-1}a\left(\frac{\partial}{\partial v}-\frac{l-1}{v}\right)F_{l-1}^{(T)}=v\left(\frac{F_{l-1}}{2l-1}-\frac{F_{l+1}}{2l+3}\right),\label{eq:Transverse}
\end{equation}
where the $J^{l}$ represent the Legendre projections of the collision
operator detailed below and $a=eE/m_{e}$. 
%We enforce the normalisation condition: 
%\begin{equation}
%4\pi\int_{0}^{\infty}F_{0}(v)v^{2}dv=1\;.
%\end{equation}
The solution of the system of equations (\ref{eq:homogenous}) and
(\ref{eq:Transverse}) provides sufficient information
to calculate the drift velocity $W$ via: 
\begin{equation}
W=\frac{4\pi}{3}\intop_{0}^{\infty}v^{3}F_{1}dv,
\end{equation}
and the characteristic energy, defined as the ratio of the transverse
diffusion coefficient $D_{T}$ to the electron mobility $\mu(=W/E)$,
via calculation of the transverse diffusion coefficient: 
\begin{equation}
D_{T}=\frac{4\pi}{3}\intop_{0}^{\infty}v^{3}F_{1}^{(T)}dv\;.
\end{equation}
%If desired, the longitudinal diffusion coefficient can be evaluated
%from the solution of (\ref{eq:longitudinal}), via
%%
%\[
%D_{L}=\frac{4\pi}{3}\intop_{0}^{\infty}v^{3}F_{1}^{(L)}dv.
%\]
%

\subsection{Collision operator for interactions in structured matter\label{sub:Collision-operator-structure}}

The collision operator appearing in (\ref{eq:BE}) describes the rate
of change of the distribution function due to interactions with the
background material. At low electron energies, where the de Broglie
wavelength of the electron is of the order of the average inter-particle
spacing $\sim N^{-1/3}$, the charged particle is best viewed as a
wave that is coherently scattered from the various scattering centres
that comprise the medium. At higher energies, the de Broglie wavelength
becomes much less than the inter-particle spacing and the effects
of coherent scattering are no longer important. In this limit, the
binary scattering approximation is recovered, although the interaction
potential is modified as discussed above. For liquid argon, the average
interparticle spacing is approximately $4.5\,\lyxmathsym{\AA}$, implying
that ``low'' energies are those less than $\sim7.4$~eV, which
is several orders of magnitude larger than the thermal energy of $\sim0.01$~eV. 

Recently, the two-term approximation of Cohen and Lekner \cite{Cohen1967}
was extended to a multi-term regime \cite{White2009,White2011}, where
the Legendre projections of the collision operator in the small mass
ratio limit were shown to be: 
\begin{equation}
J^{0}\left({\Phi_{l}}\right)=\frac{m_{e}}{Mv^{2}}\frac{d}{dv}\left\{ v\nu_{mt}(v)\left[v{\Phi_{l}}+\frac{kT}{m_{e}}\frac{d}{dv}{\Phi_{l}}\right]\right\} \label{eq:Davydov}
\end{equation}
\begin{equation}
J^{l}{\Phi_{l}}=\tilde{\nu}_{l}(v){\Phi_{l}}\qquad\text{for \ensuremath{l}\ensuremath{\ge}1},\label{eq:structurecollisionfreq}
\end{equation}
where $M$ is the mass of an argon atom, $\Phi_{l}=\left\{ F_{l},F^{(L)},F^{(T)}\right\} $
and 
\begin{equation}
	\nu_{mt}(v)=Nv2\pi\int_{0}^{\pi}\sigma^\mathrm{scr}(v,\chi)\left[1-P_{1}(\cos\chi)\right]\sin\chi 
	d\chi=Nv\sigma^\mathrm{scr}_{mt}(v),\label{eq:collisionfreq_gas}
\end{equation}
is the binary momentum transfer collision frequency in the absence
of coherent scattering effects, while

\begin{equation}
\tilde{\nu}_{l}(v)=Nv\left(2\pi\int_{0}^{\pi}\Sigma(v,\chi)\left[1-P_{l}(\cos\chi)\right]\sin\chi d\chi\right),\label{eq:collisionfreqs-1}
\end{equation}
are the structure-modified higher-order collision frequencies. The effects of 
the structure medium are encapsulated in the static structure factor $S(K)$, 
which is the Fourier transform of the pair correlator, $g(r)$, used in sections 
\ref{sec:Screening-of-the} and \ref{sec:Effective-potential-in}. The structure factor is included via the term:
\begin{equation}
	\Sigma(v,\chi)=\sigma^\mathrm{scr}(v,\chi)\; 
	S\left(\frac{2m_{e}v}{\hbar}\sin\frac{\chi}{2}\right),
\end{equation}
which represents an effective differential cross-section.
%Here $S$ is the static structure factor as discussed above.
If we represent
$\Sigma(v,\chi)$ through an expansion in terms of Legendre polynomials:
\begin{align}
\Sigma(v,\chi) & =\sum_{\lambda=0}^{\infty}\frac{2\lambda+1}{2}\Sigma_{\lambda}(v)P_{\lambda}\left(\cos\chi\right)
\end{align}
 then one can make connection with the previous calculations of the
collision matrix elements for dilute gaseous systems. The effective
partial cross-sections $\Sigma_{l}(c)$ are defined by \cite{White2009,White2011}
\begin{align}
\Sigma_{l}(v) & =2\pi\int_{-1}^{1}\Sigma(v,\chi)P_{l}(\cos\chi)d(\cos\chi)\nonumber \\
 & =\frac{1}{4\pi}\sum_{\lambda'\lambda''}\frac{(2\lambda'+1)(2\lambda''+1)}{2l+1}\sclebsq{\lambda^{\prime}}{\lambda^{\prime\prime}}l00\sigma_{\lambda'}(v)s_{\lambda''}(v),
\end{align}
 where 
\begin{align}
\sigma_{l}(v) & 
	=2\pi\int_{-1}^{1}\sigma^\mathrm{scr}(v,\chi)P_{l}(\cos\chi)d(\cos\chi)\label{partials-1}
\end{align}
and 
\begin{align}
s_{l}(v) & =\frac{1}{2}\int_{-1}^{1}S\left(\frac{2m_{e}v}{\hbar}\sin\left(\frac{\chi}{2}\right)\right)P_{l}(\cos\chi)d(\cos\chi).
\end{align}
 It then follows that

\begin{equation}
\tilde{\nu}_{l}(v)=Nv\left[\Sigma_{0}(v)-\Sigma_{l}(v)\right].\label{eq:collisionfreqs-1-2}
\end{equation}

It is sufficient for this study to consider only low energy coherent
elastic scattering processes. At higher fields, incoherent inelastic
scattering effects including excitation and ionization would need
to be considered \cite{White2009,White2011}.

\section{Results\label{sec:Results}}

Swarm experiments are a test of the particle, momentum and energy
balance in the cross-section set and the associated transport theory
or simulation. In the low-field regime considered in this manuscript,
only conservative quasi-elastic processes are operative, and hence
the ability of the calculated values of velocity drift and characteristic energy
to match the measured coefficients provide this test on momentum
and energy balance.

In the following sections we consider the calculation of the macroscopic
swarm transport properties in the gaseous and liquid environments
from the microscopic cross-sections, including screening and coherent
scattering effects as discussed above. Initially in Section \ref{sub:Electrons-in-gaseous}
we consider only the gas phase, focussing on understanding the importance
of an accurate treatment of exchange and polarization and establishing
the credibility of the initial gas-phase potential subsequently used
as input for the calculation of cross-sections for the liquid phase
environment. Transport coefficients calculated using the screened
cross-sections and associated coherent scattering effects are considered
in Section \ref{sub:Electrons-in-liquid}, where they are compared
with the available measured transport data in the liquid phase.

\subsection{Electrons in gaseous argon -- benchmarking the potential and exchange
treatment \label{sub:Electrons-in-gaseous}}

\begin{figure}
\includegraphics[width=0.6\columnwidth]{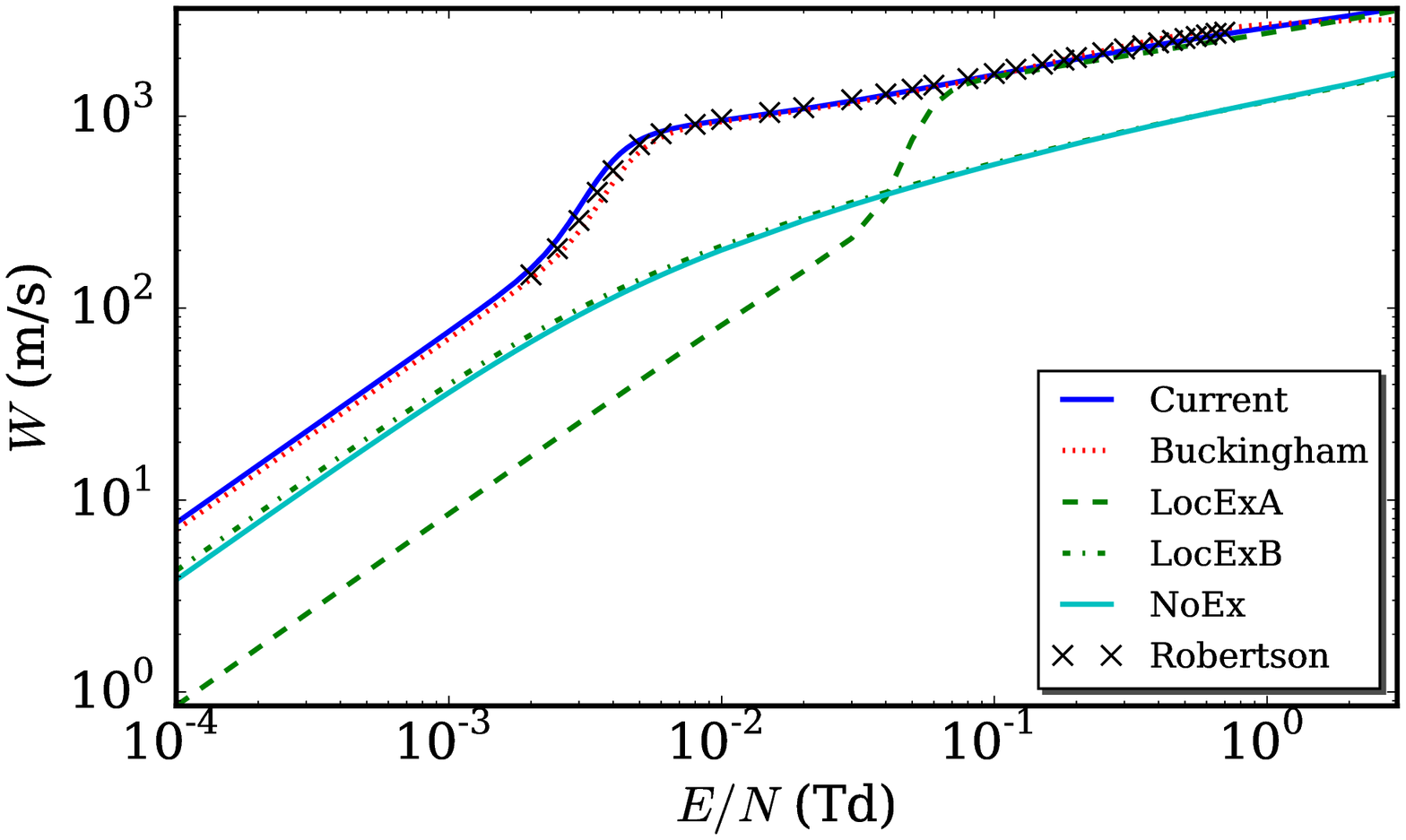}

\includegraphics[width=0.6\columnwidth]{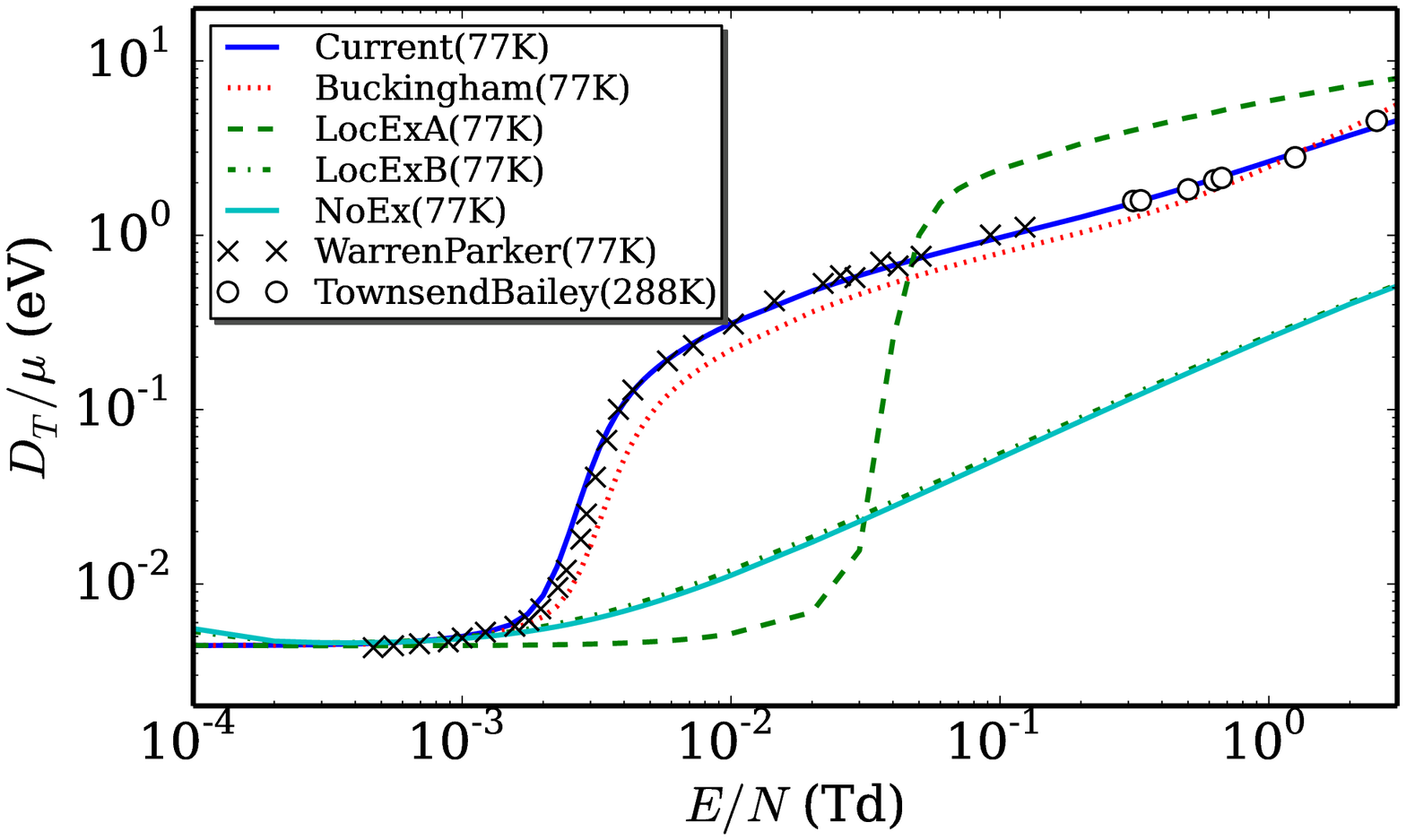}

\protect\caption{\label{fig:Drift-Gas}The drift velocity (top) and characteristic
energy (bottom) of electrons in gaseous argon, calculated using the
potentials and associated cross-sections detailed in Section \ref{sec:gas_scattering},
and compared with available experimental data (Robertson \cite{Robertson1977,LaplaceDB}
at 90~K; Warren and Parker \cite{Warren62,DuttonDB} at 77~K; Townsend
and Bailey \cite{Townsend22,DuttonDB} at 288~K). The full non-local
treatment of exchange considered here is compared to two forms of
local exchange potentials (LocExA \cite{Furness73}; LocExB \cite{Bransden76})
and to the case when the exchange interaction is neglected altogether.
The background argon gas for the calculations was fixed at 90~K for
determination of the drift velocity and 77~K for the characteristic
energy.}
\end{figure}

The calculated drift velocity and characteristic energy transport
properties using the gas-phase cross-sections detailed in Section
\ref{sec:gas_scattering} are presented in Figure \ref{fig:Drift-Gas}.
They are compared against various experimental data for this gas \cite{Robertson1977,Warren62}.
We restrict ourselves to the reduced electric fields of less than
3 Td, to ensure we are in the regime where only elastic scattering
is operative. 

Our current potential, with a non-local treatment of exchange, generally 
reproduces drift velocities to within the experimental errors. There are small 
regions of E/N (where the properties vary rapidly with E/N) where errors can be 
as large 3\% in the drift velocity and 10\% or the characteristic energy.  For 
the characteristic energy this can be outside the experimental errors in this 
region.  If the
exchange interaction is neglected in the calculation of the cross-section,
we observe that the calculated values of the transport properties
depart from the measured by an order of magnitude or more, reflecting
the qualitative disagreement in the form of the cross-sections predicted
in Figure \ref{fig:cross_sections}. Given the similarities in the
cross-sections calculated using the local exchange potential B \cite{Bransden76}
to those neglecting exchange, the calculated transport coefficients
are quite similar between the two techniques. Using the local treatment
of exchange A \cite{Furness73}, which reproduces the Ramsauer minimum
in the cross-section (although its depth, location and width disagree
quantitatively), the transport coefficients have a similar qualitative
form, however they are displaced to significantly higher fields relative
to the measured values. As expected, implementation of the Buckingham
potential as in Lekner \cite{Lekner1967}, which was tuned to reproduce
the zero-energy gas-phase cross-section, produces drift velocities
that are accurate to within 10\%, however the characteristic energies
are significantly worse than those using the current potential. 

The results shown here for argon demonstrate the validity of the electron-argon
interaction potential developed for the current study, and the necessity
for a strict non-local treatment of exchange and an accurate treatment
of polarization, in order to generate the accurate microscopic differential
cross-sections.  The small disagreement for the characteristic energy over a small range of E/N may reflect some minor limitations in the cross-section data base.

\subsection{Electrons in liquid argon\label{sub:Electrons-in-liquid}}

\begin{figure}
\includegraphics[width=0.6\columnwidth]{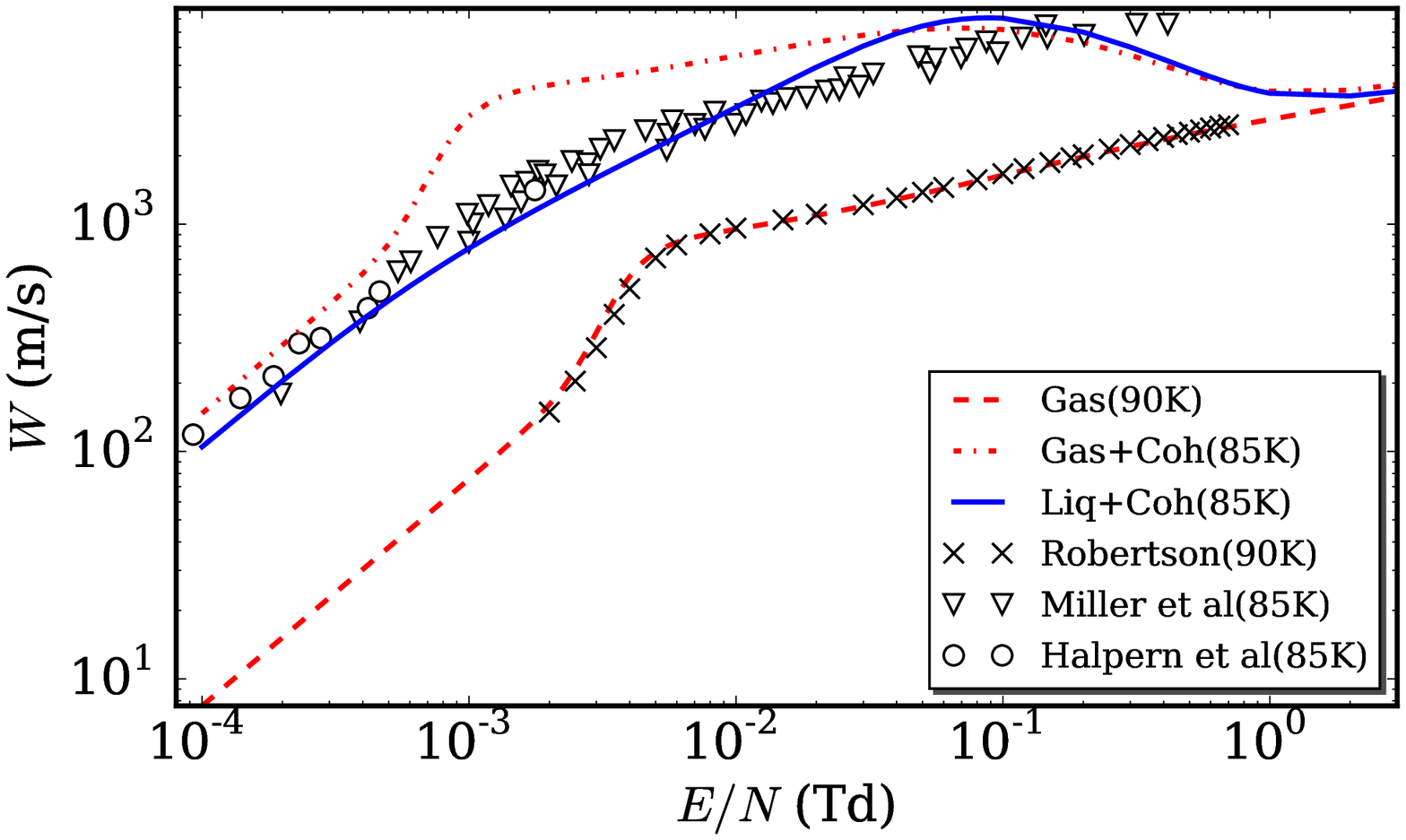}

\includegraphics[width=0.6\columnwidth]{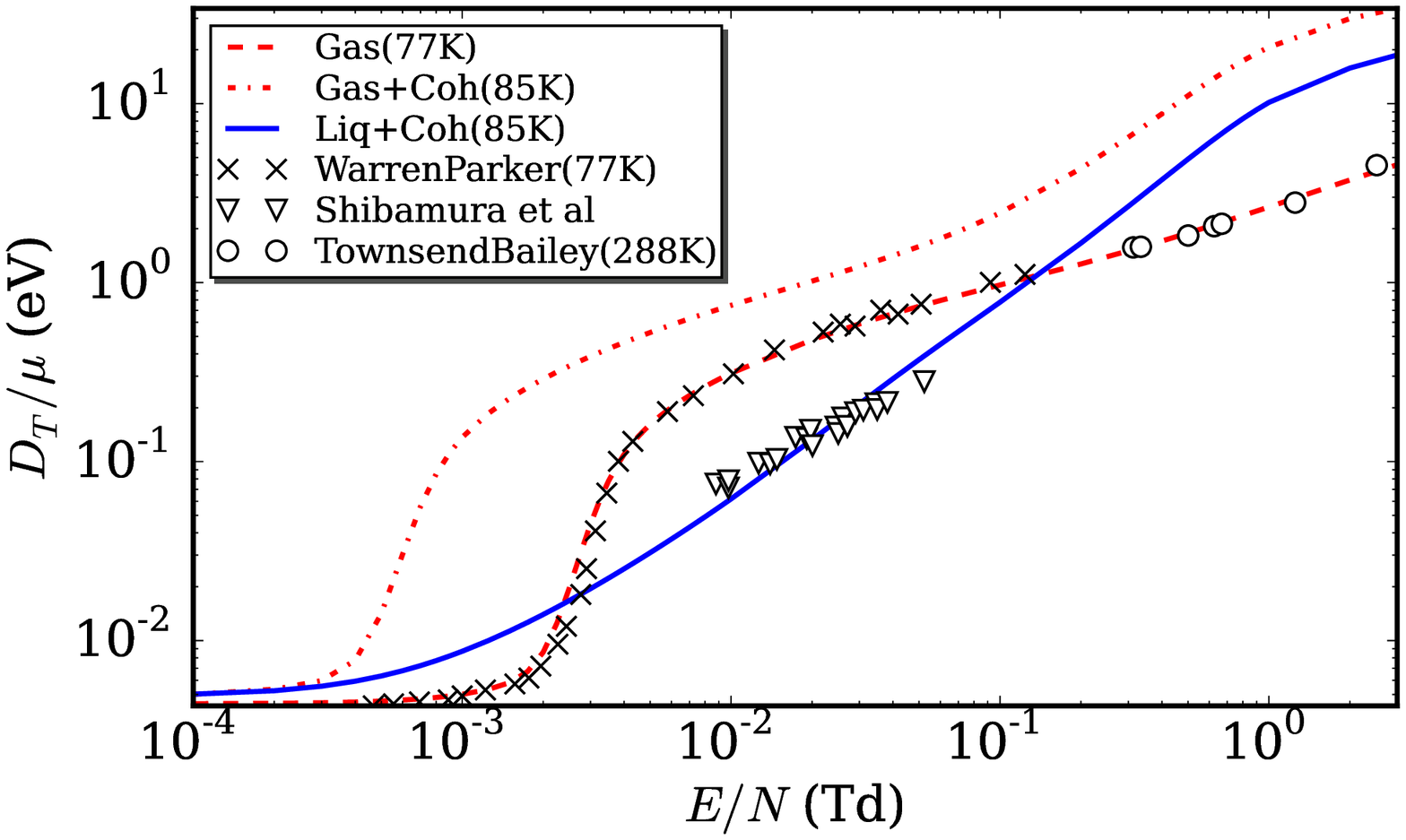}

\protect\caption{\label{fig:Liquid-Ar}Comparison of the measured drift velocities
$W$ and characteristic energies $D_{T}/\mu$ in gaseous and liquid
argon, with those calculated from the various approximations to the
cross-sections. Experimental data (Robertson \cite{Robertson1977,LaplaceDB}
at 90 K; Miller et al \cite{Miller68} at 85 K; Halpern et al \cite{Halpern67}
at 85 K; Warren and Parker \cite{Warren62,DuttonDB} at 77 K; Townsend
and Bailey \cite{Townsend22,DuttonDB} at 288~K; Shibamura et al
\cite{Shibamura1979} at an unmeasured liquid temperature). The various
approximations used are: gas-phase only cross-sections (Gas), gas-phase
cross-sections with coherent scattering (Gas+Coh), and liquid phase
cross-sections with coherent scattering effects (Liq+Coh). The results
have been calculated using the full differential cross-section and
results are converged multi-term values. Experimental uncertainties
are estimated at 2\% for Robertson and less than 15\% for Shibamura
et al.\textbf{ }}
\end{figure}

In Figure \ref{fig:Liquid-Ar}, we compare the drift velocity and
characteristic energies in both the gaseous and liquid phases. The
transport coefficients are presented as a function of the reduced
electric fields, so that the explicit density dependence has been
scaled out and we have a true comparison of the gaseous and liquid
phases. For a given reduced field, we observe that the drift velocity
in the liquid phase is enhanced by as much as an order of magnitude
over the gaseous phase in the reduced field range considered. Contrarily,
the characteristic energy in the liquid phase is reduced relative
to the gaseous phase by as much as 500\% over the range of fields
for which the data exists. Importantly, the measured data emphasizes
that transport of electrons in liquids cannot be treated by using
only the gas phase cross-sections and scaling of the density to those
of liquids.

We now assess the importance of including various physical processes
present in liquids in reproducing the measured transport coefficients.

\begin{figure}
\includegraphics[clip,width=0.6\columnwidth]{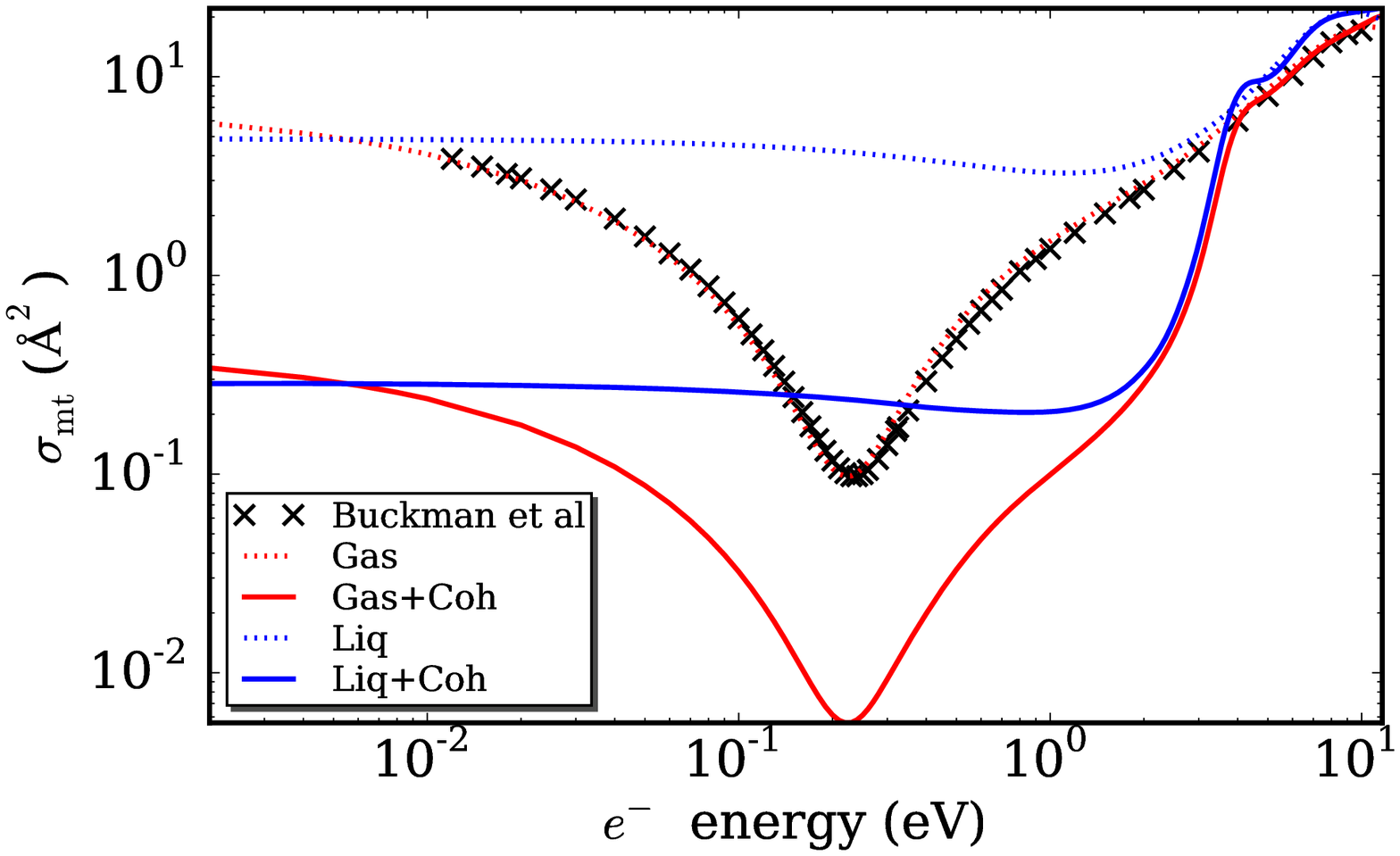}\protect\caption{\label{fig:Liquid-Ar-cross-sections} The momentum transfer cross-sections
in the gas-phase (Gas), liquid-phase (Liq) and their modifications
when coherent scattering effects are included (+Coh). The recommended
transfer cross-section of reference \cite{Buckman03} for a dilute
gas is a combination of experimental measurements and theoretical
calculations.}
\end{figure}

Firstly, we assess the importance of coherent scattering effects,
by implementing the gas-phase interaction potential and associated
cross-sections into the coherent scattering framework detailed in
Section \ref{sub:Collision-operator-structure}. The resulting cross-sections
are displayed in Figure \ref{fig:Liquid-Ar-cross-sections}. We observe
in Figure \ref{fig:Liquid-Ar} that the inclusion of only coherent
scattering effects acts to enhance both the drift velocity and the
characteristic energy. This is a reflection of the reduced momentum
transfer cross-section in Figure \ref{fig:Liquid-Ar-cross-sections}
in the regime where coherent scattering effects are operative \cite{White2011}.
Interestingly, coherent scattering produces the physical process of
negative differential conductivity (i.e. the fall of the drift velocity
with increasing electric field) which is absent from the gas-phase
calculations, as discussed elsewhere \cite{White2011}. While the
inclusion of coherent scattering effects results in a calculated drift
velocity of the same order of magnitude as the experimental data,
it does not reproduce the correct shape in the profiles, with errors
as large as 250\%. Further, the calculated characteristic energy produced
by including coherent scattering enhances the characteristic energies
relative to the gas phase which is inconsistent with the experimental
data. 

Secondly, in addition to the coherent scattering, we now include the
full liquid induced effects on the potential as detailed in Sections
\ref{sec:Screening-of-the} and \ref{sec:Effective-potential-in}.
The resulting cross-sections are displayed in Figure \ref{fig:Liquid-Ar-cross-sections},
where we emphasize that such effects act to essentially remove the
Ramsauer minimum in the cross-section. This produces an enhanced and
relatively constant cross-section in that energy regime. This is very
similar to that predicted by Atrazhev and Iakubov \cite{Atrazhev1981},
in their reduction of the Cohen and Lekner theory, which suggested
that a cross-section that is only density dependent would occur for
low impact energies. In Figure~\ref{fig:Liquid-Ar} we demonstrate
that the inclusion of both scattering potential modification and coherent
scattering produces drift velocities and characteristic energies that
are both qualitatively and quantitatively in agreement with the experimental
data. Errors in the drift velocities and characteristic energies are
significantly reduced.

\begin{figure}
\includegraphics[width=0.6\columnwidth]{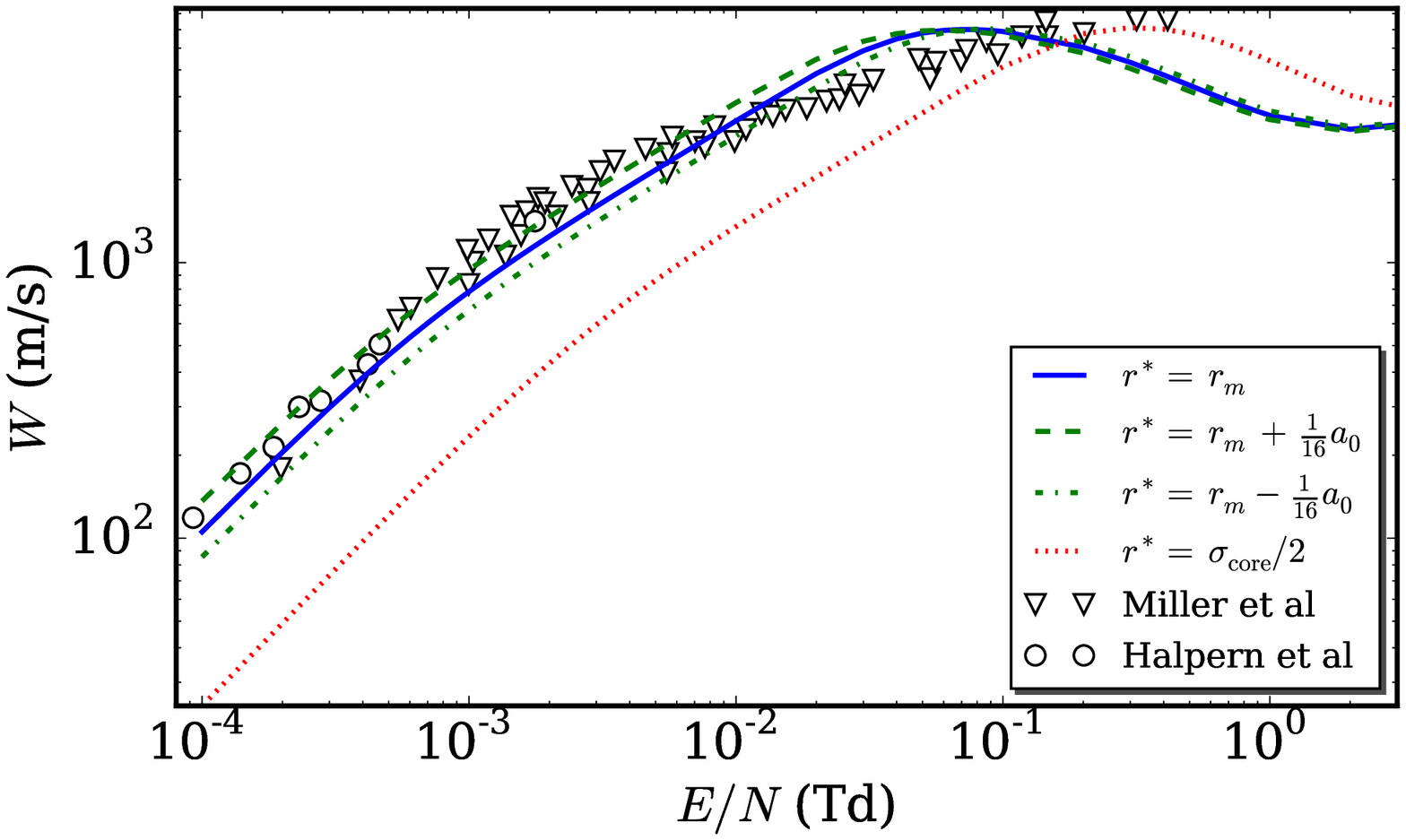}

\includegraphics[width=0.6\columnwidth]{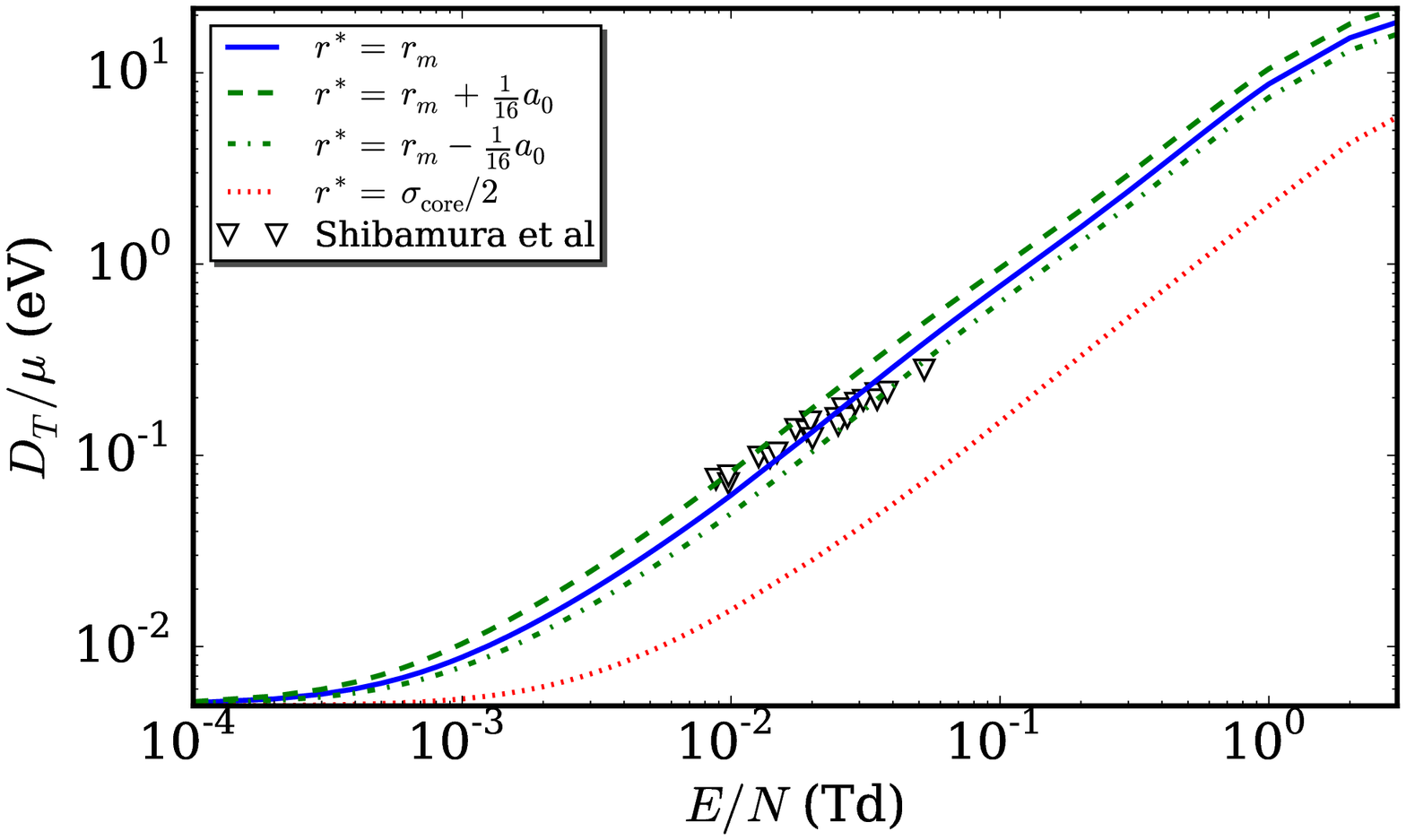}

\protect\caption{\label{fig:Impact of rm}Comparison of the calculated drift and characteristic
energy with variation in the distance $r^{*}$ at which the phase
shifts are determined. Experimental data is as detailed in Figure~\ref{fig:Liquid-Ar}.}
\end{figure}

In Figure \ref{fig:Rm_variation}, we highlighted the sensitivity
of the calculated cross-sections in the liquid phase to the value
of $r^{*}$ at which the phase shifts are determined. The macroscopic
manifestations of this sensitivity on both the drift velocity and
characteristic energy is presented in Figure \ref{fig:Impact of rm}.
Slight modifications of $r^{*}$ by \textbf{$a_{0}/16$} from the
preferred value of $r^{*}=r_{m}$ emphasize the sensitivity of the
transport coefficients to this value. The choice of $r^{*}=\sigma_{\mathrm{core}}/2$
produces results that are essentially translated to higher reduced
electric fields. Importantly, these results indicate that the value
of $r_{m}$ may be energy dependent. One could possibly tune the value
of $r_{m}$ to match the experimental data, however we have strived
to eliminate adjustable parameters in our formalism. One may also
look at using an alternative scheme that is energy-dependent for choosing
the value of $r_{m}$, e.g. including contributions from the exchange
terms $\lineW_{P,\mathrm{eff}}$ and $\lineW_{Q,\mathrm{eff}}$.

\subsection{Impact of scattering anisotropy and the two-term approximation}

We conclude this study by considering the impact of the anisotropy
in both the scattering cross-sections and the velocity distribution
function on the calculated transport properties.

\begin{figure}
\includegraphics[clip,width=0.9\columnwidth]{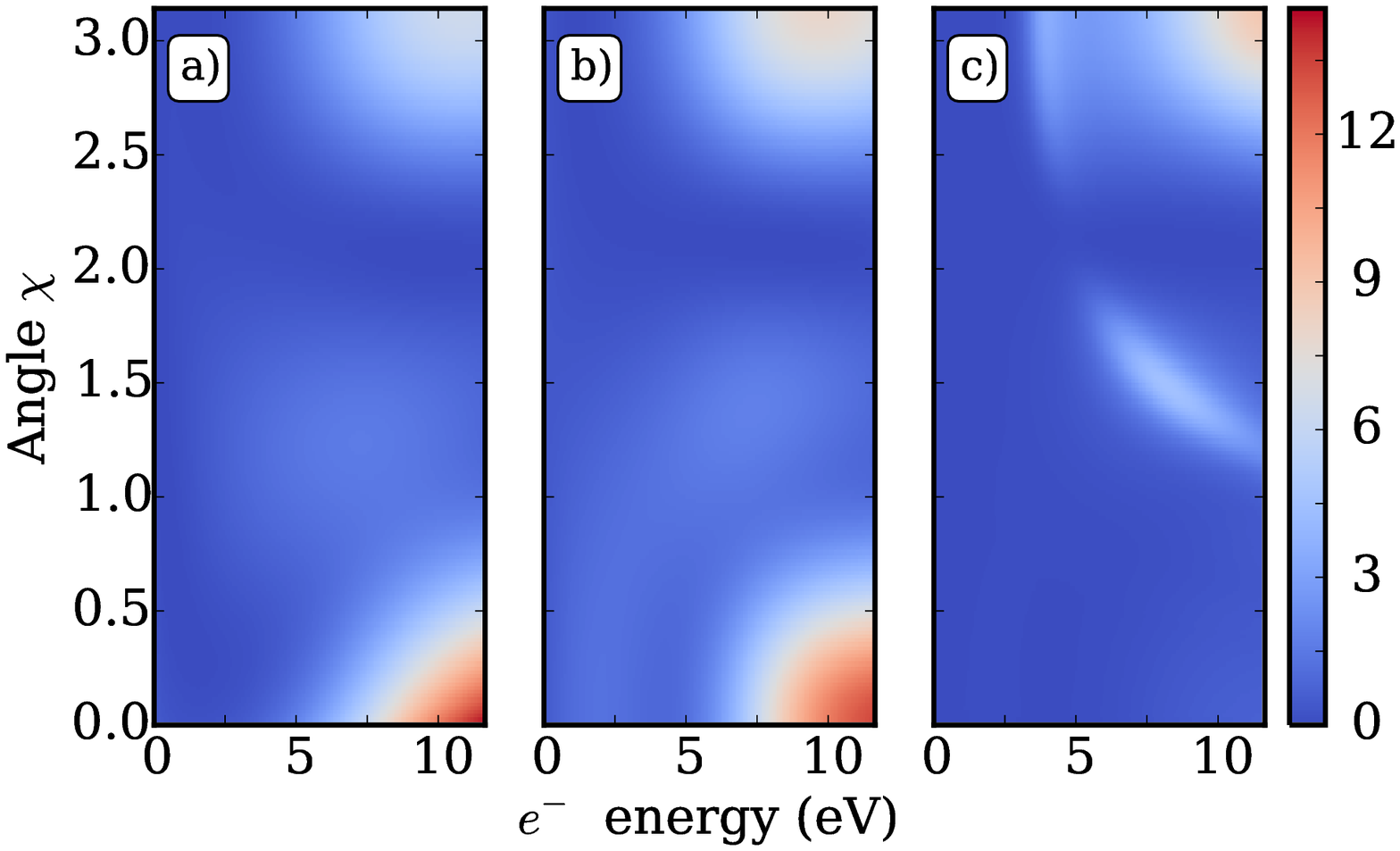}
\includegraphics[clip,width=0.9\columnwidth]{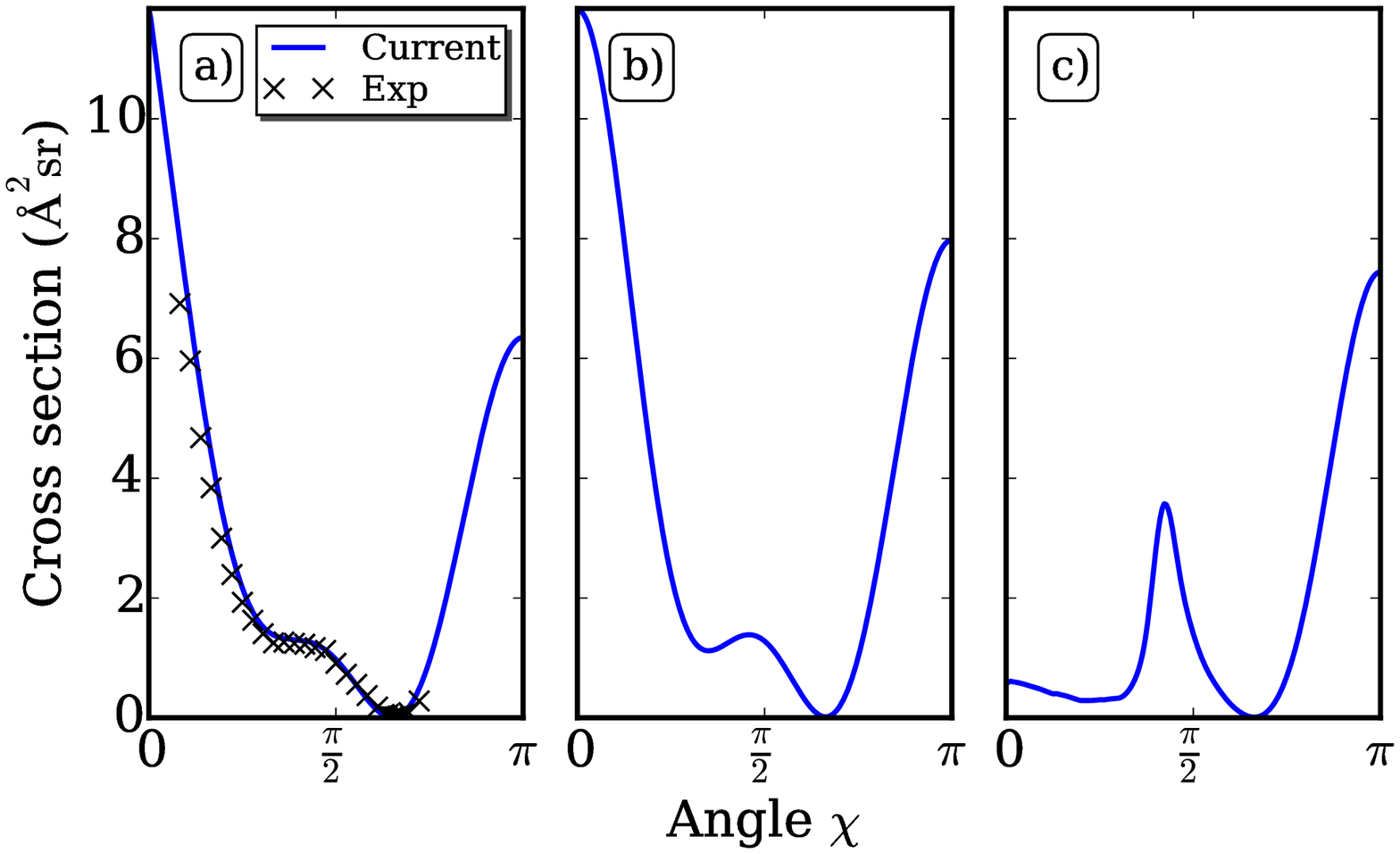}
\protect\caption{\label{fig:DifferentialCrossSections} Top: Differential cross-sections
	in square angstroms per steradian for electrons in Ar for a) dilute gas 
	phase, $\sigma(\epsilon,\chi)$ b)
	effective liquid phase including screening effects, 
	$\sigma^\mathrm{scr}(\epsilon,\chi)$ and c) liquid
phase cross-section including coherent scattering effects $\Sigma(\epsilon,\chi)$. Bottom: Differential cross-sections taken at 10 eV for the same cases.  The experimental data for the gas phase is taken from Gibson et al. \cite{Gibsonetal96}.}
\end{figure}

In Figure \ref{fig:DifferentialCrossSections} we display the differential
cross-sections for the gas phase phase and for the liquid modified
differential cross-sections, highlighting the impact of coherent scattering
effects. For the dilute gas phase, we observe at low energies that
the differential cross-sections are small and essentially isotropic.
As we move to higher energies, the differential cross-section begins
to demonstrate an increased magnitude and also enhanced anisotropy,
with peaks in the forward and back-scattering directions. This is confirmed by agreement with the experimental data of Gibson et al.  \cite{Gibsonetal96}. When we
account for liquid effects in the scattering potential, we observe
that similar qualitative structures are present in the resulting differential
cross-section, with slightly more structure than for the dilute gas
phase. When the liquid phase differential cross-section is combined
with the structure factor accounting for coherent scattering effects,
the resulting differential cross-section $\Sigma(\epsilon,\chi)$
takes on a completely different qualitative structure. The forward
peak in the differential cross-section is removed, with suppression
of the cross-section at low energies and low scattering angles. The
backscattering peak in the differential cross-section at high energies
remains unaffected, while subpeaks in the differential cross-section
are enhanced by the coherent scattering effects. 

\begin{figure}
\includegraphics[clip,width=0.6\columnwidth]{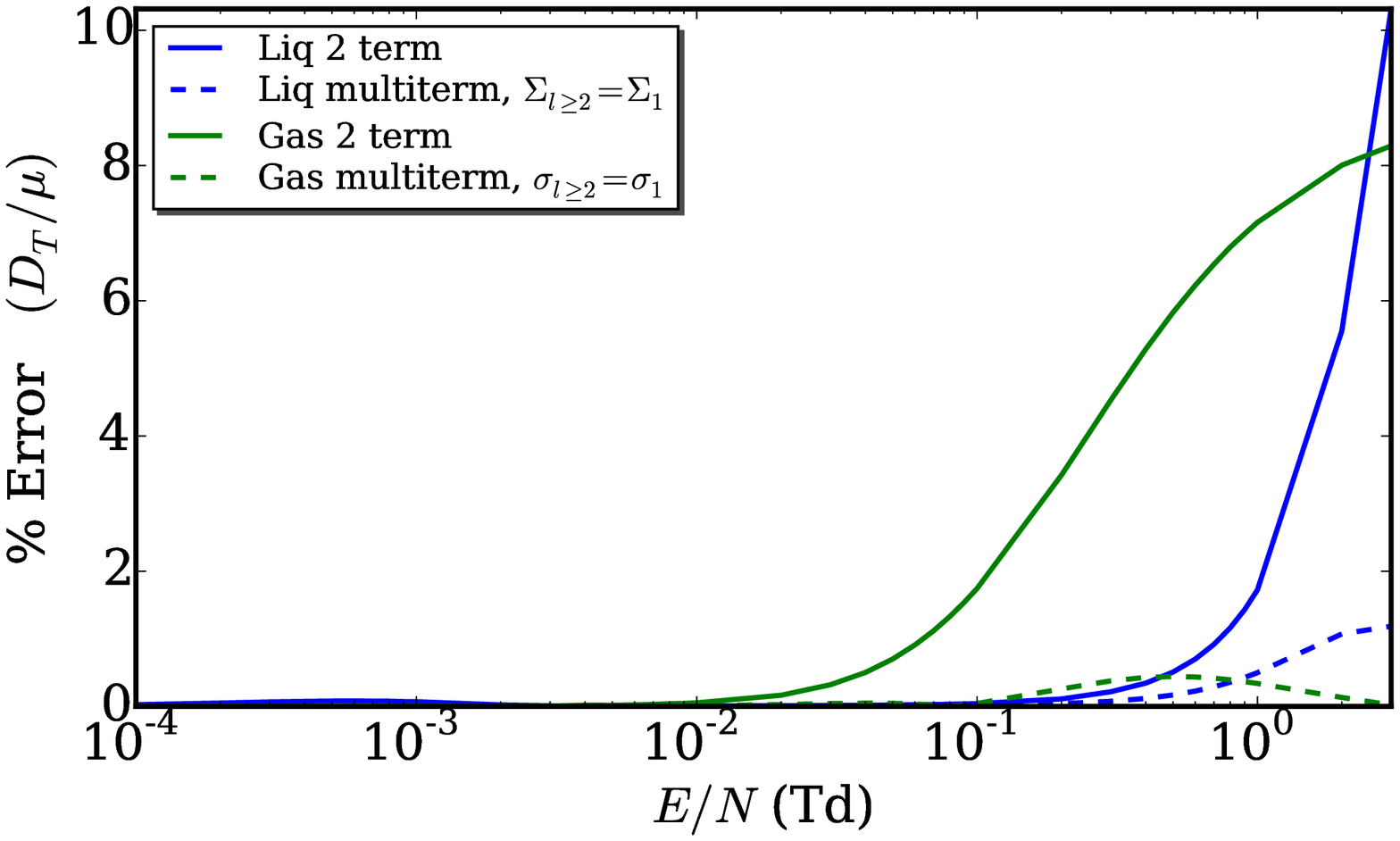}\protect\caption{\label{fig:Two-term approximation} Percentage differences between
the two-term and multi-term values of the characteristic energy for
the gas and liquid phases using the full differentail cross-sections
(solid lines), and percentage differences between the multi-term results
with using only the momentum transfer cross-section and the full differential
cross-section (dashed lines). All percentages are relative to the
converged multi-term result using the full differential cross-section. }
\end{figure}

The degree of anisotropy in the distribution function is evidenced
by an enhanced value of $l_{\mathrm{max}}$ required in the spherical
harmonic expansions (\ref{sphericalharmonic-1})
to achieve convergence in the velocity distribution or transport properties.
In Figure \ref{fig:Two-term approximation}, we display the error
in the two-term approximation ($l_{\mathrm{max}}=1)$ and the converged
multi-term result. In the gas and liquid phases we see that the two-term
approximation is sufficient to ensure accuracy to within 0.5\% in
the drift velocity, however errors as large at 10\% are present in
the characteristic energy. This indicates a failure of the two-term
approximation for the evaluation of the characteristic energy. Similar
findings in the gas-phase were found by Brennan and Ness \cite{Brennan1992b}.
Theories that have used the two-term approximation to iteratively
adjust cross-sections may produce cross-sections that are inconsistent
with a multi-term framework. 

In Figure \ref{fig:Two-term approximation} we also consider the impact
of anisotropic scattering on the validity of the two-term approximation.
The two-term approximation can only sample the momentum transfer cross-section.
Higher-order spherical harmonic components of the distribution function
in expansions (\ref{sphericalharmonic-1})
are coupled to, and hence sample, higher-order coefficients in the
expansion of the differential cross-section (see e.g. equation (\ref{partials-1})).
In Figure~\ref{fig:Two-term approximation} we highlight the differences,
using dashed lines, between the multi-term approximation using only
the momentum transfer cross-section (i.e. we assume $\sigma_{l\geq2}=\sigma_{1}$)
and those where the full differential cross-section is considered.
The differences are less than 1\% (usually less than 0.1\%) indicating
the distribution function is not sufficiently anisotropic to couple
in higher-order partial cross-sections. Equivalently, anisotropy in
the differential cross-sections has only a minimal impact on the anisotropy
in the velocity distribution function.

\section{Conclusions}

We have extended the approach of Lekner and Cohen \cite{Cohen1967,Lekner1967},
overcoming some of its limitations, to calculate the effective cross-sections
and transport properties of electrons in liquid argon. For the first
time an accurate multipole polarisability in the electron-atom potential,
and a fully non-local treatment of exchange were included in the calculation
of liquid phase cross-sections using the full machinery of the Dirac-Fock
scattering equations. The accuracy of the potential implemented and
associated cross-sections calculated was confirmed by comparison with
experiment in the gas-phase, and the importance of a fully non-local
treatment of exchange was demonstrated. The result calls into question
cross-sections (gas, liquid or clusters) which assume a local treatment
of exchange. Sensitivity to the radial cut-off for the electron-atom
potential was presented, and while the maximum in the potential was
shown to be a suitable choice, enhanced accuracy may be achieved with
an energy dependent choice of the cutoff. 

The calculation of the drift velocity and characteristic energies
were performed for the first time using a multi-term solution of Boltzmann's
equation accounting for coherent scattering. The full anisotropy of
the liquid-phase differential cross-section was considered including
anisotropy arising from both the interaction and from the structure
factor. The multi-term framework enabled an assessment of the sensitivity
to this anisotropy in the differential cross-section and in the velocity
distribution function. While the two-term approximation was found
to be sufficient for accuracies to within 1\% for the drift velocity,
errors of the order of 10\% or more were found in the characteristic
energy. The latter was found to be the dominant contribution to the
differences in the two and multi term results. It was found that both
coherent scattering and screening of the electron-atom potential are
required to reproduce the measured transport coefficient values. We
emphasize that there are no free parameters in the current theory
and its implementation, and hence the high level of agreement between
the calculated and measured transport coefficients yields confidence
that the essential physics has been captured in the theory. 

\section{Acknowledgements}

The authors acknowledge the financial assistance of the Australian Research 
Council (ARC) through its Discovery and Centres of Excellence programs.

\bibliographystyle{apsrev}
%\bibliography{library,library_extra}

\end{document}